\documentclass[lettersize,journal]{IEEEtran}
\usepackage{cite}

\usepackage[pdftex]{graphicx}
\usepackage{amsmath,amssymb,amsfonts}
\usepackage{algorithmic}
\usepackage[ruled]{algorithm2e}
\usepackage{multirow}
\usepackage{tabularx}
\usepackage{bm}
\usepackage{textcomp}
\usepackage{subfigure}
\usepackage{xcolor,soul,framed}
\graphicspath{{fig/}}
\usepackage{balance}
\usepackage{bm}

\usepackage{xcolor}
\usepackage{xpatch}

\begin{document}
%
\title{UAV Trajectory Planning for AoI-Minimal Data Collection in UAV-Aided IoT Networks by Transformer}
	\author{Botao~Zhu,~\IEEEmembership{}
		Ebrahim~Bedeer,~\IEEEmembership{Member,~IEEE,}
		~Ha H. Nguyen,~\IEEEmembership{Senior Member,~IEEE},
Robert~Barton,~\IEEEmembership{Member,~IEEE}, and Zhen Gao,~\IEEEmembership{Member,~IEEE}
		\thanks{B. Zhu, E. Bedeer, and H. H. Nguyen are with the Department of Electrical and Computer Engineering, University of Saskatchewan, Saskatoon, Canada S7N5A9. Emails: \{botao.zhu, e.bedeer, ha.nguyen\}@usask.ca.}
		\thanks{R. Barton is with Cisco Systems Inc. Email: robbarto@cisco.com.}
        \thanks{Z. Gao is with Beijing Institute of Technology, Beijing, China. Email: gaozhen16@bit.edu.cn.}
 		\thanks{This work was supported by NSERC/Cisco Industrial Research Chair in Low-Power Wireless Access for Sensor Networks.}
	}
		
	\maketitle
	
	\begin{abstract}
	Maintaining freshness of data collection in Internet-of-Things (IoT) networks has attracted increasing attention. By taking into account age-of-information (AoI), we investigate the trajectory planning problem of an unmanned aerial vehicle (UAV) that is used to aid a cluster-based IoT network. An optimization problem is formulated to minimize the total AoI of the collected data by the UAV from the ground IoT network. Since the total AoI of the IoT network depends on the flight time of the UAV and the data collection time at hovering points, we jointly optimize the selection of hovering points and the visiting order to these points. We exploit the state-of-the-art transformer and the weighted A*, which is a path search algorithm, to design a machine learning algorithm to solve the formulated problem. The whole UAV-IoT system is fed into the encoder network of the proposed algorithm, and the algorithm's decoder network outputs the visiting order to ground clusters. Then, the weighted A* is used to find the hovering point for each cluster in the ground IoT network. Simulation results show that the trained model by the proposed algorithm has a good generalization ability to generate solutions for IoT networks with different numbers of ground clusters, without the need to retrain the model. Furthermore, results show that our proposed algorithm can find better UAV trajectories with the minimum total AoI when compared to other algorithms.
\end{abstract}
	
		
	\IEEEpeerreviewmaketitle
	
	\section{Introduction}\label{SecI}

	\IEEEPARstart{T}he use of unmanned aerial vehicles (UAVs) has attracted a lot of attention from academia and industry. Given their high maneuvering capability and mobility, UAVs can be used as wireless relays or mobile base stations to provide reliable communications and better coverage for ground devices \cite{9446488}. Thanks to these advantages, UAVs can be flexibly deployed to provide fast and reliable network access in different applications, such as disasters \cite{8660516}, surveillance \cite{9309248}, monitoring \cite{9201322}, to name a few.

	Since UAVs can fly close to the ground devices and provide low-altitude air-to-ground communication links with them, UAVs can be deployed to hover the area of interest to collect data from ground Internet-of-Things (IoT) networks. By doing so, UAV-aided data collection can save the energy of devices in traditional IoT networks, thus extending their lifetime \cite{9507262}. However, maintaining the freshness of the collected information is an important issue in time-sensitive IoT applications, such as environmental monitoring and safety protection. In these applications, the generated data needs to be sent to the destination as soon as possible. Outdated information can lead to incorrect control and even cause major disasters \cite{zhang2020age}. Therefore, it is essential to ensure the freshness of the data received at the destination. To measure the freshness of information, the age of information (AoI) as a new performance metric was proposed in \cite{6195689}. In a nutshell, AoI describes the amount of time elapsed since the generation of the most recent data update. AoI-based data collection can guarantee information freshness in IoT networks, which is quite different from traditional delay-based and throughput-based metrics \cite{9151993}. As such, it has attracted increasing attention.
	
	Due to the importance of AoI, a number of studies have been carried out on AoI-oriented data collection in UAV-assisted wireless networks. In \cite{8406973}, the authors aimed to minimize the average AoI of the system by optimizing the trajectory of the UAV in a UAV-aided data collection system. In \cite{9286911}, the authors optimized the trajectory of the UAV to minimize the maximal AoI and the average AoI of sensors. In \cite{8756751}, the authors assumed the UAV supports three modes to collect data and jointly optimize the trajectory and data collection modes of the UAV to minimize the average AoI of all ground nodes. In \cite{8570843}, the UAV trajectory, energy, and service time allocation were jointly optimized by an iterative algorithm in order to minimize the overall peak AoI of the system. The authors in \cite{9285214} developed an energy-efficient navigation policy for the UAV to improve data freshness of the IoT network. In order to minimize the weighted sum of AoI, the authors in \cite{9162896} jointly optimized the flight trajectory of the UAV and the transmission scheduling of sensors. From the above discussion, it can be seen that AoI-oriented data collection problems in the UAV-assisted IoT network are typically related to UAV's trajectory design.


	When collecting data in the UAV-assisted IoT network, if the UAV is dispatched to visit every ground IoT device, the energy consumption of the UAV will increase because of the increased UAV trajectory. Hence, to reduce the energy consumption of the UAV, clusters-based  model have been extensively investigated in UAV-assisted wireless networks. For instance, in \cite{8515012}, to gather compressive data measurements, the authors divide the sensor network into multiple clusters. In each cluster, all nodes build a forwarding tree based on compressive data gathering to send data to the cluster head (CH). The UAV then traverses all CHs to collect the aggregated data. The authors jointly optimized the UAV trajectory, CH selection, and forward tree construction to minimize the total transmit power in the network. In \cite{9416816}, the authors consider a pre-clustered network where a UAV equipped with multiple antennas  communicates with multiple ground users simultaneously, in a given time slot, using space division multiple access. The authors jointly optimized the time slot allocation and the UAV hovering time to minimize the overall energy consumption. In \cite{9148990}, the authors considered a UAV-enabled data collection system for massive machine-type communications (mMTC) where machine-type communication devices (MTCDs) are divided into several clusters. A UAV visits each hovering position which corresponds to a MTCD cluster and sequentially collects data from each MTCD in the corresponding cluster. They formulated a problem of minimizing the total energy consumption of the system. In our previous work \cite{9507262}, we considered using a UAV to collect data from a clustered IoT network, where the hovering points of the UAV are determined by the unknown CHs location. In other words,  in \cite{9507262}, we jointly select the CHs and their visiting order to minimize the total energy consumption. In this paper, we examine the scenario where the UAV collects data from a group of clusters and the UAV only interact with the CHs. The problem of interest in this paper is to jointly optimize the UAV's hovering points and trajectory to achieve the minimal AoI data collection in a cluster-based IoT network. The optimization problem is formulated as a traveling salesman problem (TSP) with neighborhoods (TSPN), which is extremely challenging because it includes a continuous problem (optimization of hovering points) and a combinatorial problem (optimization of visiting order).

        The hovering points of the UAV and the visiting order to these hovering points have a great impact on the flying time of the UAV and data collection time, which directly influence the total AoI of collected data. There have been some works on solving the TSPN efficiently.
        In \cite{a6010084}, the Dubins TSPN was converted to a generalized TSP (GTSP) by using the sampling-based roadmap method, and then to an asymmetric TSP that can be addressed by the Lin-Kernighan heuristic algorithm. To handle the continuous optimization problem of waypoints within each circular neighborhood, the authors in \cite{pvenivcka2017dubins} proposed a discretization scheme that equidistantly samples possible locations along the circular border of the interest neighborhood to determine the locations of the waypoints. In this paper, in order to reduce the computational complexity for solving the joint optimization of the UAV's hovering points and trajectory to achieve the minimal AoI data collection in a cluster-based IoT network, we transform the formulated continuous optimization TSPN into a GTSP by borrowing the sampling-based idea. The transformed GTSP is a combinatorial optimization problem that can be solved using traditional methods, such as exact algorithms, approximate algorithms, or heuristic algorithms. However, these traditional algorithms may not achieve a good balance between optimality and computational complexity. Thus, by considering optimality, computational complexity, and generality, we shall develop a machine learning-based algorithm to solve the transformed GTSP, i.e., the UAV's trajectory  design problem.

        Machine learning has been explored as a promising technique for solving UAV's trajectory planning problems in UAV-assisted IoT networks. To minimize the weighted sum-AoI in a UAV-assisted network, the authors in \cite{9013924} applied deep reinforcement learning (DRL) to optimize the UAV's trajectory using a deep Q network (DQN) and an artificial neural network (ANN). In \cite{8752017}, the authors utilized Q-learning to optimize AoI-optimal UAV path by considering the deadline constraints of data in the UAV-aided sensing network. In \cite{9374448}, the authors jointly optimized the UAV's trajectory and  scheduling of the status update packets to minimize the normalized weighted sum of AoI in a UAV-assisted wireless network. Specifically, they used ANN, DQN, and long short-term memory (LSTM) to develop a DRL algorithm for learning the UAV trajectory in large-scale networks. Different from these works, we employ the state-of-the-art transformer and the weighted A* search method to design a UAV trajectory planning algorithm for AoI-oriented data collection.

		Transformer was originally proposed by Google as a sequence-to-sequence model to deal with machine translation problem \cite{vaswani2017attention}. It has achieved great success in many areas of artificial intelligence in the past four years, such as computer vision, audio processing, document summarization, and document generation. Some researchers also attempt to use transformer and its variants to tackle combinatorial problems, such as the TSP. In \cite{deudon2018learning}, the cities in the TSP were encoded by a transformer and decoded sequentially through a query consisting of the last three cities in the partial tour. The used transformer was trained by reinforcement learning. In \cite{kool2018attention}, the authors also used the transformer architecture as the encoder network and the decoder network outputs the result sequentially based on the embeddings from the encoder and the outputs generated at previous steps. The encoder and decoder networks were trained using a reinforce algorithm with a deterministic greedy baseline. The authors in \cite{9393606} proposed a transformer-based framework to automatically learn improved heuristics on two representative routing problems: the TSP and capacitated vehicle routing problem (CVRP). In \cite{bresson2021transformer}, the authors used the standard transformer architecture to tackle TSP and achieve an improved performance over recent learned heuristics. Inspired by the success of employing transformer in solving various problems of route planning, we propose the transformer-weighted-A* (TWA*) algorithm in this paper for solving our formulated GTSP combinatorial optimization problem. Although the Ptr-A* algorithm proposed in our previous work \cite{9507262} achieves good performance in solving the GTSP, the TWA* algorithm has the following two important advantages over the Ptr-A* algorithm. First, TWA* does not relay on past hidden states like Ptr-A*, and thus, avoids losing past information. Second, TWA* has the ability of parallel computation which makes it faster than Ptr-A*.
		
		The main contributions of this paper are summarized as follows:
	\begin{enumerate}
		\item We propose an AoI-oriented data collection model in a cluster-based IoT network and formulate a total AoI-minimal trajectory planning problem where the hovering points of the UAV and the visiting order to these points are jointly optimized.
		\item We view the formulated problem as a ``machine translation'' problem where the ``source language'' is the whole UAV-IoT network and the ``target language'' is the UAV trajectory with the minimal total AoI. The state-of-the-art TWA* is employed to solve the formulated problem. The parameters of the proposed algorithm are trained by reinforcement learning that only needs the reward calculation.
		\item The learned policy by the proposed algorithm generalizes well on different sizes of problem instances. In other words, the trained model by the proposed algorithm can automatically find a trajectory with the minimal total AoI for new problem instances, without retraining the model.
		\item Extensive simulations are conducted to evaluate the performance of the proposed algorithm. Results show that the proposed algorithm achieves significant performance gain in maintaining data freshness while reducing computation time when compared with other baseline algorithms.				
	\end{enumerate}	
	
	The rest of this paper is organized as follows. Section \ref{SecII} introduces the system model and presents the formulated problem. Section \ref{SecIII} develops the proposed algorithm. Section \ref{SecIV} provides simulation results. Finally, Section \ref{SecV} concludes the paper.

\section{System Model and Problem Formulation} \label{SecII}
	We consider a UAV-assisted IoT network that consists of one rotary-wing UAV, one ground base station (BS) located at $b_0$, and $M$ clusters of ground sensor nodes. Specifically, each cluster $m$, $m=1,\dots,M$, has one CH, located as $b_m$, and $N_m$ ordinary sensor nodes, located at as $B_m = \{b_m^{(1)}, \dots, b_m^{(N_m)}\}$. The ground IoT network performs some sensing tasks in the surrounding area where the ordinary sensor nodes are responsible for sampling data and forwarding the collected data to their corresponding CHs. The UAV is dispatched from the start hovering point $c_0$ which is directly above $b_0$ to visit $M$ mission hovering points $\{c_1, \dots, c_m, \dots, c_M\}$ by a pre-designed trajectory for data collection, and then flies back to  $c_0$ after completing the data collection task. Each hovering point corresponds to one ground cluster and its position will be determined by the proposed algorithm. The three-dimensional (3D) Cartesian coordinates system is considered to define positions of hovering points and all CHs. The coordinate of the $m$-th hovering point is denoted by ${c}_{m} = (x_{c_{m}},y_{c_{m}},H) \in \mathbb{R}^3$, where $H$ is the flight height of the UAV, whereas the location of the corresponding ground CH is given by ${b}_{m} = (x_{b_{m}},y_{b_{m}},0)\in \mathbb{R}^3$.
	
\begin{figure}[t]
		\centering
		\includegraphics[width=3in]{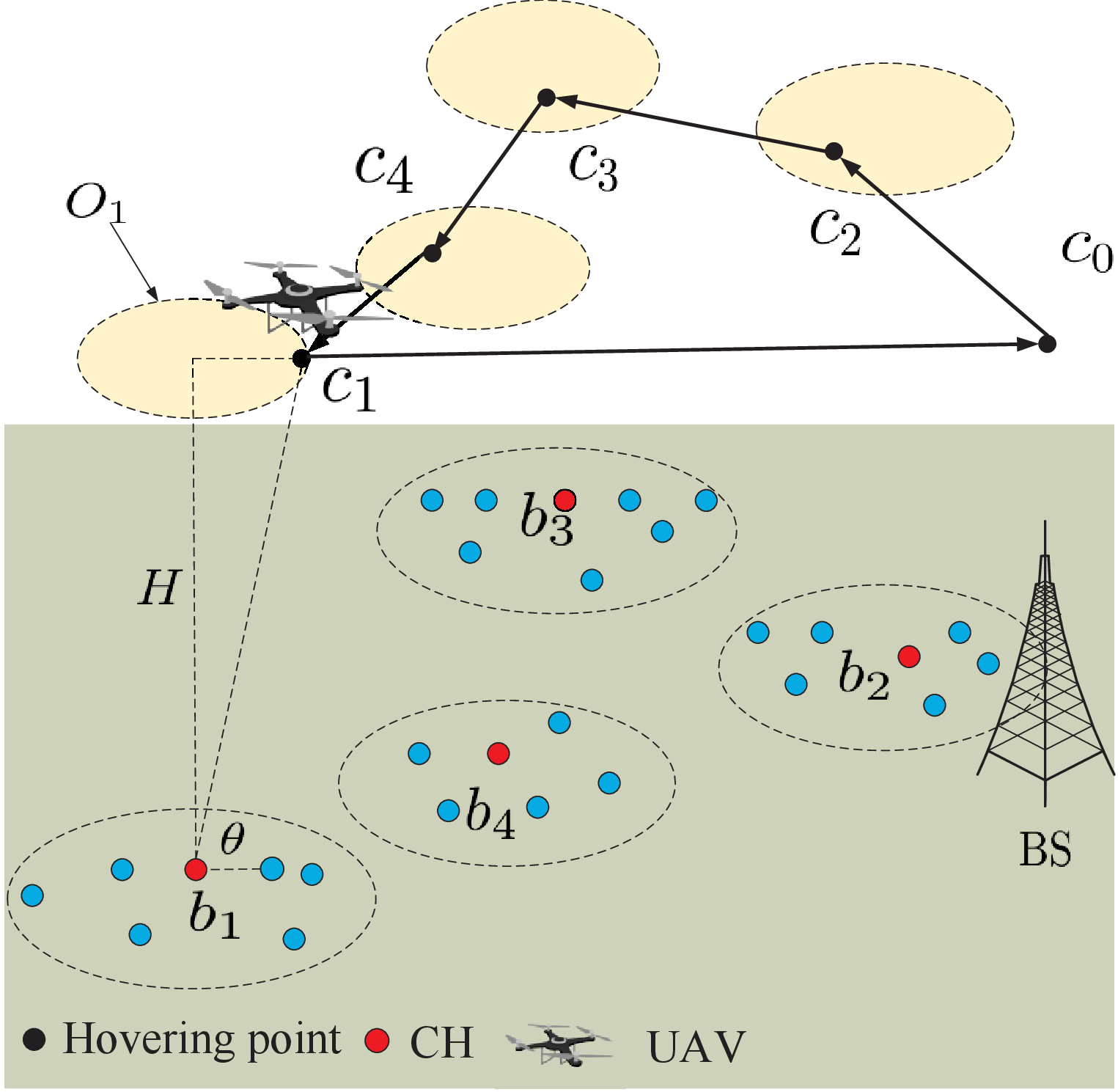}
		\caption{System model of a UAV-assisted IoT network.}
		\label{systemmodel}
	\end{figure}

	We assume that the rotary-wing UAV supports a flying-hovering mode without considering acceleration-deceleration, i.e., it flies to the hovering points with a fixed speed $v_{\text{UAV}}$ and hovers at these points with static status to collect data from ground CHs. We illustrate the UAV-assisted data collecting process in Fig. \ref{systemmodel}. The UAV takes off from ${c}_0$, determines the position of the hovering point ${c}_2$ that will be visited first and arrives at it. The UAV repeats this procedure until data collection of all clusters is completed, and flies back to ${c}_{0}$. Hence, the final trajectory of the UAV in this example is $\{c_0,c_2,c_3,c_4,c_1,c_0\}$.
	
\subsection{Data Collection Model}
   	
When the UAV arrives at $c_{m}$, it sends a beacon message to wake up the corresponding CH ${b}_{m}$ from its sleep mode. The beacon message includes the type of sensor nodes to be activated in response to the beacon, the data collection height of the UAV, a threshold to limit the number of sensor nodes in the CH (if necessary), and a trailer that has error detection capabilities.
Then, ${b}_{m}$ switches to its active mode and informs its member nodes in the same cluster to sample and send their sampled data sequentially according to the pre-allocated equal-length time slots using time-division multiplexing (TDM) protocol to avoid collision. We consider the \emph{generate-at-will} model \cite{8000687} as the data sampling model for all ordinary sensor nodes, by which nodes can generate information updates at any time. Specifically, we assume that each node can generate an update message of size $L_{\text{data}}$ only in its allocated time slot to eliminate the waiting time. Also, each message has a time stamp, which is the start of each time slot. The length of a time slot is denoted as $\tau$ seconds. After the CH located at $b_{m}$ finishes collecting data from its member nodes, it will forward the collected data to the UAV. For ease of analysis, the wake-up time of nodes, including CHs and all ordinary nodes, and the information sampling time of each node are assumed negligible as compared to the data collection time. Thus, the data collection time of the UAV at each hovering point mainly consists of two parts: the data transmission time from ordinary nodes to their CHs and the time consumed for forwarding the collected data from CHs to the UAV.
	
We consider both the line-of-sight (LoS) and non-line-of-sight (NLoS) links to design the ground-to-air communication when the UAV hovers at mission hovering points. The LoS link probability is related to environment, elevation angle, and transmission distance, which can be expressed as \cite{al2014optimal}
	\begin{equation}
	    \label{plos} P_{c_{m}}^{({\rm LoS})} = \frac{1}{1 + \beta \exp{\left(- \widetilde{\beta}  \left(\theta_{c_{m}} -\beta\right)\right)}},
	\end{equation}
	where $\beta$ and $\widetilde{\beta}$ are constants determined by the environment,
	$\theta_{c_{m}} = \arctan{(H/R_{(c_{m},b_{m})})}$ is the elevation angle between  $b_{m}$ and the UAV when it hovers at $c_{m}$, $R_{(c_{m},b_{m})} = \sqrt{\left(x_{c_{m}}-x_{b_{m}}\right)^2+ \left( y_{c_{m}}-y_{b_{m}}\right)^2}$ is the horizontal distance between the CH $b_{m}$ and the hovering point.  Correspondingly, the probability of NLoS is given by $P^{({\rm NLoS})}_{c_{m}}=1-P_{c_{m}}^{({\rm LoS})}$.
	In addition, the path loss models of LoS and NLoS between the CH $b_{m}$ and the UAV follow \cite{8708295}
	  \begin{align}
	    \label{llos} L_{c_{m}}^{({\rm LoS})} & = 20\log_{10}\left(\frac{4 \pi f_c d_{(c_{m},b_{m})}}{v_{\text{light}}} \right) + \xi_{\text{LoS}}, \\
	    L_{c_{m}}^{({\rm NLoS})} & =  20\log_{10}\left(\frac{4 \pi f_c d_{(c_{m},b_{m})}}{v_{\text{light}}} \right) + \xi_{\text{NLoS}},
	\end{align}
	where $f_c$ is the carrier frequency, $v_{\text{light}}$ is the speed of light, $d_{(c_{m},b_{m})} = \sqrt{H^2+R^2_{{(c_{m},b_{m})}}}$ is the distance between the UAV and the CH $b_{m}$, $\xi_{\text{LoS}}$ and $\xi_{\text{NLoS}}\,(\xi_{\text{LoS}}<\xi_{\text{NLoS}})$ are the excessive path losses in LoS and NLoS links, respectively. We consider the average path loss to describe the link from the ground CH to the UAV, which can be expressed as

	\begin{equation}
	   \label{averageloss}
	    \overline{L}_{c_{m}} = P_{c_{m}}^{({\rm LoS})}L_{c_{m}}^{({\rm LoS})} + P_{c_{m}}^{({\rm NLoS})}L_{c_{m}}^{({\rm NLoS})}.
	\end{equation}
	To avoid the interference among CHs, we assume that only one CH can transmit data to the UAV at any given time. Hence, the average available transmission rate in bits per second (bps) from CH $b_{m}$ to the UAV can be expressed as $r_{c_{m}} = B_{\text{width}} \log_2\left (1 + \gamma_{c_{m}} \right)$,
	where $B_{\text{width}}$ is the channel bandwidth in hertz (Hz), $\gamma_{c_{m}} = {P_{\text{CH}}}/\left({\sigma^2 10^{\overline{L}_{c_{m}}/10}}\right)$ is the signal-to-noise ratio (SNR) of the transmission link, $\sigma^2$ is the noise power at the UAV, and $P_{\text{CH}}$ is the transmission power of the CH. Regarding the transmission quality, we set a SNR threshold $\gamma_\text{th}$  and the transmission is considered successful if the SNR is greater than the threshold. Thus, the SNR constraint at the UAV receiver is given as
	\begin{equation}
	\label{snr}
	    \gamma_{c_{m}} \ge \gamma_{\text{th}}.
	\end{equation}
	
	\emph{Lemma 1:} Given the fixed flight height $H$,  $c_{m}$ should be located in a horizontal disk region centered at the position that directly above $b_{m}$ and having the radius $R^*$ which can guarantee that the UAV successfully receives data. When $R_{(c_{m},b_{m})}=R^*$, the received SNR of the UAV at $c_{m}$ is equal to $\gamma_{\text{th}}$.
	
	\emph{Proof}: See Appendix A.
	
	Based on \emph{Lemma 1}, we formally define a hovering disk region for each mission hovering point (excluding the start point $c_0$) as
	\begin{equation}
	\label{diskregion}
	    O_{m} = \{c_{m} :||c_{m}-b{'}_{m}||=R_{(c_{m},b_{m})} \leq R^*\}
	\end{equation}
where $b{'}_{m}=\left(x_{b_{m}}, y_{b_{m}}, H\right) \in \mathbb{R}^3$ is the center of the disk $O_{m}$, and $R^*$ is the radius to maintain a pre-defined quality-of-service, which can be found numerically. As long as the UAV enters a hovering disk region, it can collect data from the corresponding ground CH. The total data collection time of the UAV at $c_{m} \in O_{m}$ (or its hovering time) can be simply written as
	\begin{equation}
	\label{transsiontime}
	    T^{({\rm hov})}_{c_{m}} = N_{{m}} \tau + \frac{N_{{m}} L_{\text{data}}}{r_{c_{m}}}
	\end{equation}
	where the first term in the right hand side is the time consumed for transmitting data from ordinary nodes to their corresponding CH $b_{m}$, and the second term is the data transmission time from $b_{m}$ to the UAV. Therefore, the energy consumption of propulsion-related and communication-related activities of the UAV while hovering at $c_{m}$ is expressed as
	\begin{equation}
	\label{consumption_hovering}
	    E_{c_{m}} = P_{\text{hov}}T^{({\rm hov})}_{c_{m}} +  P_{\text{com}}\frac{N_{m} L_{\text{data}}}{r_{c_{m}}}
	\end{equation}
	where $P_{\text{hov}}$ and $P_{\text{com}}$ are the UAV's powers for hovering and communication, respectively. After finishing the data collection task, ${b}_m$ switches to the sleep model for saving energy. The UAV continues to select the next hovering point and executes the same processes to collect the sensed data from the corresponding ground cluster.

	\subsection{UAV's Mobility Model}
	Without loss of generality, the flight trajectory of the UAV can be seen as a permutation of the visiting order to $M$ mission hovering points, with the start point being $c_0$, i.e., $\bm{c} = \{c_0,c_1,\dots,c_M\}$. The set of all possible permutations is denoted as $\bm{\Phi}$ with the size of $M!$. We represent one of the permutations as $\bm{\pi} = \{\pi(0), \dots, \pi(M+1)\}$ and express the ordered hovering points as $\bm{c}_{\bm{\pi}} = \{c_{\pi(0)}, c_{\pi(1)}, \dots, c_{\pi(M)}, c_{\pi(M+1)} \}$, where $c_{\pi(t)}, t = 0, \dots, M+1$, is the hovering point that is visited at step $t$ in the trajectory, and $c_{\pi(0)}=c_{\pi(M+1)} =c_0$. For ease of understanding, if the hovering point $c_m$ is visited at step $t$, its corresponding cluster of ground ordinary nodes ($B_m$) and the number of ordinary nodes ($N_m$) are redefined as $B_{\pi(t)}$ and $N_{\pi(t)}$, respectively.
		
	After finishing data collection at $c_{\pi(t)}$ with the hovering model, the UAV horizontally flies to the next hovering point $c_{\pi(t+1)}$ along the line segment connecting $c_{\pi(t)}$ and $c_{\pi(t+1)}$. The flying time of the UAV during this period is given by

	\begin{equation}
	\label{flighttime}
	    T^{({\rm fly})}_{(c_{\pi(t)}, c_{\pi(t+1)})} = \frac{||c_{\pi(t)}-c_{\pi(t+1)}||}{v_{\text{UAV}}}
	\end{equation}
	where $||c_{\pi(t)}-c_{\pi(t+1)}||$ is the Euclidean distance between $c_{\pi(t)}$ and $c_{\pi(t+1)}$.

	Following \cite{8663615}, the propulsion power consumption of the UAV for horizontal movement is the function of speed $v_{\text{UAV}}$ and given by
	\begin{align}
	    \label{pf}
	    P_{\text{mov}}(v_{\text{UAV}})
	    =  &P_0 \left(1+\frac{3v_{\text{UAV}}^2}{U_{\text{tip}}^2}\right) +  P_1\left(\left(1+\frac{v_{\text{UAV}}^4}{4v_0^4}\right)^{1/2}-\frac{v_{\text{UAV}}^2}{2v_0^2} \right)^{1/2} \nonumber \\ & + \frac{1}{2}d_0\rho s_0 \delta v_{\text{UAV}}^3
	\end{align}
	where $P_0$ and $P_1$ represent, respectively, the blade profile power and induced power in the hovering state, $U_{\text{tip}}$ is the tip speed of the rotor blade of the UAV, $v_0$ is the mean rotor induced velocity in the hovering state, $d_0$ denotes the fuselage drag ratio, $s_0$ represents the rotor solidity, $\rho$ is the density of air, and $\delta$ denotes the area of the rotor disk.
	According to the analysis in \cite{8663615}, the power consumption $P_{\text{mov}}(v_{\text{UAV}})$ firstly decreases and then increases with the increasing value of the speed $v_{\text{UAV}}$.
	The energy consumption in the UAV's flight from $c_{\pi(t)}$ to $c_{\pi(t+1)}$ is computed as
	\begin{equation}
	\label{consumption_flying}
	    E_{(c_{\pi(t)}, c_{\pi(t+1)})}=P_{\text{mov}}(v_{\text{UAV}}) T^{({\rm fly})}_{(c_{\pi(t)}, c_{\pi(t+1)})}.
	\end{equation}
    In the hovering state, the power consumption of the UAV can be obtained by substituting $v_{\text{UAV}}=0$ into (\ref{pf}), $P_{\text{hov}} = P_0 + P_1$, which is a constant value.

	\subsection{Age of Information Model in a UAV-IoT System}
    \begin{figure}[!t]
		\centering
		\includegraphics[width=0.9\linewidth]{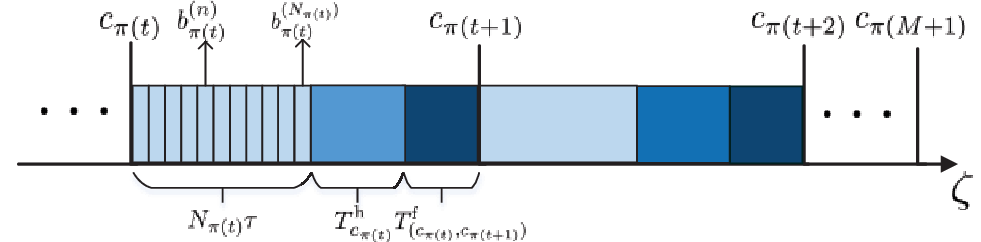}
		\caption{The time sequence of data collection in the considered UAV-IoT system.}
		\label{aoi}
	\end{figure}
	We use the AoI metric to measure the freshness of information. According to the definition of AoI in \cite{tong2019uav},  the AoI of a packet collected from node $b^{(n)}_{\pi(t)}$ in the $\pi(t)$-th visited cluster at time $\zeta$ is defined as
	 	\begin{equation}
	        A^{(n)}_{\pi(t)}(\zeta) =  \left(\zeta - u^{(n)}_{\pi(t)}(\zeta)\right)^+
	    \end{equation}
	where $u^{(n)}_{\pi(t)}(\zeta)$ is the instant at which the packet is generated, and $(x)^+=\max\{0,x\}$. When $\zeta < u^{(n)}_{\pi(t)}(\zeta)$, we define $A^{(n)}_{\pi(t)}(\zeta) = 0$. This is because the packet of node $b^{(n)}_{\pi(t)}$ has not been sampled. It is evident that the AoI of a packet will increase with time. In the considered UAV-IoT system, the BS is seen as the observer, thus, the AoI of a data packet can be seen as the amount of time elapsed from the instant at which the packet is generated to the instant at which the UAV flies back with the collected data to the BS.
	
	For ease of analysis, for any ordinary node $b_{\pi(t)}^{(n)}, n=1, \dots, N_{\pi(t)}$ in the $\pi(t)$-th visited cluster, the AoI of its packet can be simply divided into two components. The first component is the time needed for the CH of its associated cluster to collect data from $b_{\pi(t)}^{(n)}$ and other nodes whose data have not been gathered (i.e., nodes $b_{\pi(t)}^{(n+1)},\dots, b_{\pi(t)}^{(N_{\pi(t)})}$) and forward the collected data to the UAV. The second component is the time consumed by the UAV to carry the packet of $b_{\pi(t)}^{(n)}$ to the end point $c_{\pi(M+1)}$. Specifically, this period includes the flight time of the UAV to unvisited ground clusters and the data collection time in these clusters. For example, after completing the data collection at $c_{\pi(t)}$,  the UAV will fly to the next hovering point $c_{\pi(t+1)}$ and gather information from the corresponding cluster. During this period, the AoI of the packet of $b_{\pi(t)}^{(n)}$ increases with time, which is the sum of the flight time $T_{(c_{\pi(t)}, c_{\pi(t+1)})}$ from $c_{\pi(t)}$ to $c_{\pi(t+1)}$ and data collection time $T_{c_{\pi(t+1)}}$ at hovering point $c_{\pi(t+1)}$. Then, the UAV performs the same process to unvisited clusters until it returns to the end point. The time sequence of data collection in the UAV-IoT system is illustrated in Fig. \ref{aoi}. Mathematically, the total AoI of the packet generated by $b_{\pi(t)}^{(n)}$ in the UAV-IoT system is given as
	\begin{align}
	    \label{aoie}
	    A_{\pi(t)}^{(n)} =\, & \underbrace{ \left(N_{\pi(t)}-\left(n-1\right)\right) \tau + \frac{N_{\pi(t)}L_{\text{data}}}{r_{c_{\pi(t)}}}}_{\text{first component}} \nonumber \\ &+ \underbrace{\sum_{g=t}^{M-1} \left(T^{({\rm fly})}_{(c_{\pi(g)}, c_{\pi(g+1)})} + T^{({\rm hov})}_{c_{\pi(g+1)}}\right) +  T^{({\rm fly})}_{(c_{\pi(M)}, c_{\pi(M+1)})}}_{\text{second component}}
	\end{align}
	 which can be further simplified as
		\begin{equation}
		\label{aoiee}
	    A_{\pi(t)}^{(n)} = \sum_{g=t}^{M} \left(T^{({\rm hov})}_{c_{\pi(g)}} + T^{({\rm fly})}_{(c_{\pi(g)}, c_{\pi(g+1)})} \right)
	    - (n-1) \tau.
	\end{equation}
	For packets of nodes in the same cluster, we have
	\begin{equation}
	\label{intraaoi}
	    A_{\pi(t)}^{(1)} >  A_{\pi(t)}^{(2)} \dots >  A_{\pi(t)}^{(N_{\pi(t)})}.
	\end{equation}
	On the other hand, the AoIs of packets in different clusters should satisfy
	\begin{equation}
	\label{clusteraoi}
	     A_{\pi(1)}^{(n)}>  A_{\pi(2)}^{(n)}> \dots >  A_{\pi(M)}^{(n)}.
	\end{equation}

	\subsection{Problem Formulation}
	
	The total AoI of all ordinary nodes in the network can be computed as
	\begin{align}
	\label{averageaoi}
	    \overline{A} = \sum_{t=1}^{M}\sum_{n=1}^{N_{\pi(t)}}A_{\pi(t)}^{(n)} \nonumber
	      = & \sum_{t=1}^{M}\sum_{n=1}^{N_{\pi(t)}} \sum_{g=t}^{M} \left(T^{({\rm hov})}_{c_{\pi(g)}} + T^{({\rm fly})}_{(c_{\pi(g)}, c_{\pi(g+1)})} \right) \\
	    &- \sum_{t=1}^{M}\sum_{n=1}^{N_{\pi(t)}} (n-1) \tau.
	\end{align}
	
	According to (\ref{averageaoi}), the total AoI is expressed as a weighted sum of the flight time of the UAV and the data collection time at each hovering point, which is determined by the locations of hovering points $\bm{c}$, the visiting order to these hovering points $\bm{\pi}$. It is evident that the hovering points of the UAV and its trajectory have a strong impact on the total AoI of data. If the position of any hovering point $c_m$ is close to the center of the disk region $O_m$, a high data transmission rate can be achieved. As a result, the data transmission time from CHs to the UAV can be reduced, even though the UAV may have a longer flight trajectory, and hence the flight time. Conversely, if the UAV is located near the boundary of the disk region, the length of the UAV's trajectory might be reduced, but it will result in a lower data transmission rate, and hence increased data transmission time.
		
	Our objective is to jointly find the hovering point from each disk and plan the visiting order to these hovering points for the UAV to minimize the total AoI of data in the considered UAV-IoT system. The optimization problem is expressed as follows:
	 \begin{alignat}{2}
	       \mathcal{P}_1: \min_{\substack {\bm{c},\bm{\pi}}} \quad &  \overline{A} \left(\bm{c},\bm{\pi}\right), \tag{18a}\\
          \mbox{s.t.}\quad
          & \label{positionc} \bm{\pi}\in \bm{\Phi},& \tag{18b}\\
          & (\ref{snr}), (\ref{transsiontime}), (\ref{flighttime}), (\ref{intraaoi}), \: \text{and} \: (\ref{clusteraoi}). &\quad& \nonumber
    \end{alignat}
    
    Constraint (\ref{positionc}) is the trajectory constraint. The SNR constraint is given in (\ref{snr}), and (\ref{transsiontime}) is the data collection constraint. The flight time constraint is expressed as (\ref{flighttime}). AoI constraints are (\ref{intraaoi}) and (\ref{clusteraoi}). It is evident that the formulated problem $\mathcal{P}_1$ is a TSPN \cite{dumitrescu2003approximation}, which combines the determination of hovering points at each disk with the problem of trajectory planning of the UAV. The traditional TSPN problem involves finding a minimum-cost tour (i.e., the total length of the tour is minimum) that travels each region exactly once for a collection of compact regions before returning to the initial departure point\cite{yuan2007optimal}. However, our formulated problem not only considers the traveling cost but also the cost spent at each hovering point. The problem $\mathcal{P}_1$ is extremely challenging because it is composed of a continuous problem (optimization of hovering points $\bm{c}$) and a combinatorial problem (optimization of visiting order $\bm{\pi}$). Given a set of hovering points $\bm{c}$, the optimization of $\bm{\pi}$ can be viewed as the TSP, which can be normally be solved quite effectively by some dedicated TSP solvers, such as Concorde \cite{Concorde}, etc. However, the optimization of hovering points $\bm{c}$ consists of an infinite number of variables, which is infeasible to be solved optimally. To reduce computational time, we leverage the sampling approach that samples finite discrete sets of hovering points from a continuous state space to transform the continuous TSPN in $\mathcal{P}_1$ into the GTSP. Specifically, each disk $O_m$ is equally partitioned into $L_{\text{sub}} \times L_{\text{sub}}$ sub-regions and the center of each sub-region is selected as the possible hovering point. For some marginal sub-regions with non-square shape, we choose the centers of their actual areas. Hence, we can obtain a cluster $G_m$ of sampling points with the size of $L_{\text{sub}}^2$ from $O_m$. As a result, our objective is changed to jointly select hovering points from $M$ clusters of sampled hovering points and plan the UAV's trajectory to visit selected hovering points exactly once to minimize the total AoI. Using the sampling approach, the formulated problem $\mathcal{P}_1$ is converted to
      \begin{alignat}{2}
	       \mathcal{P}_2: \min_{\substack {\bm{c},\bm{\pi}}} \quad &  \overline{A} \left(\bm{c},\bm{\pi}\right), \tag{19a}\\
          \mbox{s.t.}\quad
          & \label{positionconstraint} c_m \in G_m, G_m \in O_m, m \in \{1,\dots,M\},& \tag{19b}\\
          & (\ref{snr}), (\ref{transsiontime}), (\ref{flighttime}), (\ref{intraaoi}), (\ref{clusteraoi}), \: \text{and} \: (\text{\ref{positionc}}). &\quad& \nonumber
    \end{alignat}
Obviously, the formulated problem $\mathcal{P}_2$ is a combinatorial optimization problem, and hence, NP-hard.  There are two traditional methods to handle combinatorial problems: exact algorithms and heuristic algorithms. Exact algorithms can find optimal solutions, but they will become intractable when the size of problems grows. Heuristic algorithms' complexity is polynomial and they commonly find sub-optimal solutions. In contrast, we cast the proposed GTSP as a sequence-to-sequence problem where the source sequence is a set of clusters of hovering points and CHs and the target sequence is a set of selected hovering points and the visiting order to these points. We adopt the transformer, the weighted A*, and reinforcement learning to efficiently solve this problem.

\section{Transformer-Weighted A* Algorithm}	\label{SecIII}

Because the UAV needs to sequentially collect data from each ground cluster in the IoT network, we view the problem of the total AoI-minimal trajectory planning as a ``machine translation'' problem that is common in natural language processing. The whole UAV-IoT network as the ``source language'' is translated into the ``target language'', i.e., the UAV trajectory, by using our proposed TWA* algorithm. The TWA* algorithm is composed of an encoder network, a decoder network, and the weighted A* search algorithm which can effectively find the trajectory policy from hidden patterns behind a large number of training datasets.

   \subsection{Encoder}
    The role of the encoder network is to take the UAV-IoT network represented as an input sequence and map it into an abstract representation that is the learned information. The input sequence includes the start point of the UAV, each and every CH, number of nodes in each ground cluster, and all sampling points from each hovering disk. Specifically, we define $\bm{h}_0^{({\rm in})}=c_0 \in \mathbb{R}^{3}$, and $\bm{h}_m^{({\rm in})} = \left(G_m, b_m,N_m\right) \in \mathbb{R}^{3(L_{\text{sub}}^2 + 1) + 1}, m \in \{1, \dots, M\}$, where the cluster $G_m$ of sampling points is represented as a $3L_{\text{sub}}^2$-dimensional vector as it includes $L_{\text{sub}}^2$ points with 3D Cartesian coordinates, the CH $b_m$ is a 3-dimensional vector, and the number of nodes $N_m$ is a constant. Hence, the input can be expressed as $\bm{H}^{({\rm in})} = \left(\bm{h}_0^{({\rm in})}; \bm{h}_1^{({\rm in})}; \dots; \bm{h}_M^{({\rm in})} \right)$.
     \begin{figure*}[!t]
		\centering
		\includegraphics[width=0.85\linewidth]{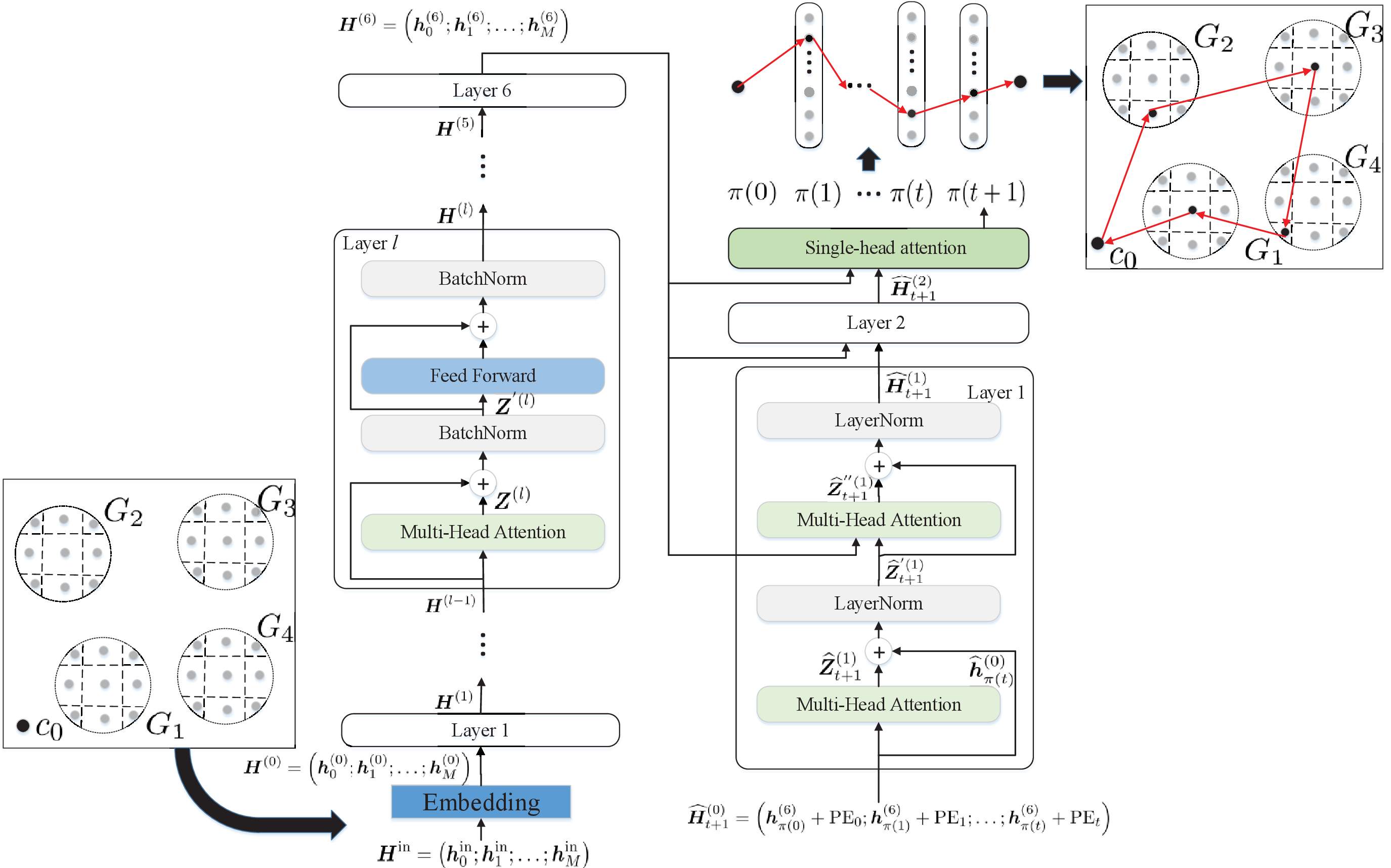}
		\caption{The proposed algorithm framework.}
		\label{transformer}
     \end{figure*}
    The encoder network used in this paper is the standard transformer encoder with one embedding layer and six identical encoder layers as in \cite{bresson2021transformer}. Each encoder layer is composed of one multi-head self attention sub-layer and one point-wise feed-forward network sub-layer. Each sub-layer adds a residual connection and layer normalization. The embedding layer is to map each element of input to the $d_{\text{em}}$-dimensional vector space by a learnable linear projection. Specifically, to enable the model to distinguish the start point of the UAV from clusters, we separately utilize different parameters to compute the embeddings of the start point and the other clusters as follows:
    \begin{equation}
    \setcounter{equation}{20}
    \bm{h}^{(0)}_m = \left\{ \begin{array}{ll}
     \bm{W}_{0} \bm{h}_m^{({\rm in})} + \bm{W}_{b_0}, & m = 0 \\
     \bm{W}_{1} \bm{h}_m^{({\rm in})} + \bm{W}_{b}, & m=1,\dots,M
      \end{array} \right.
   \end{equation}
   where $\bm{W}_{0} \in \mathbb{R}^{d_{\text{em}} \times 3}$, $\bm{W}_{1} \in \mathbb{R}^{d_{\text{em}} \times (3(L_{\text{sub}}^2 + 1) + 1)}$, $\bm{W}_{b_{0}} \in \mathbb{R}^{d_{\text{em}}}$, and $\bm{W}_{b} \in \mathbb{R}^{d_{\text{em}}}$ are learnable parameters.
   Then, the embeddings $\bm{H}^{(0)} = \left(\bm{h}^{(0)}_0; \bm{h}^{(0)}_1; \dots; \bm{h}^{(0)}_M \right) \in \mathbb{R}^{(M+1) \times d_{\text{em}}}$ are fed into the encoder layers. Note that we do not consider the positional decoding used in the original transformer in \cite{vaswani2017attention} because the order of the input sequence is irrelevant to the GTSP.

  The attention layer in each encoder layer uses the multi-head self-attention mechanism with 8 heads to jointly attend to information from different representation subspaces at different positions. The 8 heads perform the attention calculation in parallel and their results are merged to produce an input for the next step. In the encoder layer $l, l = 1,\dots, 6$, the output of self-attention on the $h$-th head, $h = 1,\dots, 8$, is computed as
\begin{align}
\label{singleattention}
    \bm{Z}_{h}^{(l)} &= \text{Attention} (\bm{Q}_{h}^{(l)}, {\bm{K}_{h}^{(l)}}, \bm{V}_{h}^{(l)}) \nonumber \\ & = \text{softmax} \left( \frac{\bm{Q}_{h}^{(l)} {\bm{K}_{h}^{(l)}}^{T}}{\sqrt{d_{\text{v}}}}\right)\bm{V}_{h}^{(l)}
\end{align}
where $d_{\text{v}}$ is used for scaling the dot products, $\bm{Q}_{h}^{(l)}\in \mathbb{R}^{(M+1) \times d_{\text{v}}}$, $\bm{K}_{h}^{(l)}\in \mathbb{R}^{(M+1) \times d_{\text{v}}}$, and $\bm{V}_{h}^{(l)}\in \mathbb{R}^{(M+1) \times d_{\text{v}}}$ are matrices query, key, and value for the $h$-th head, respectively. They can be created by projecting the input query $\bm{Q}^{(l)}$, key $\bm{K}^{(l)}$, and value $\bm{V}^{(l)}$ of multi-head self attention with three learnable weight matrices $\bm{W}_{h}^{Q(l)} \in \mathbb{R}^{d_{\text{em}}\times d_{\text{v}}}$, $\bm{W}_h^{K(l)} \in \mathbb{R}^{d_{\text{em}}\times d_{\text{v}}}$, and $\bm{W}_h^{V(l)} \in \mathbb{R}^{d_{\text{em}}\times d_{\text{v}}}$, respectively, as follows
\begin{equation}
    \bm{Q}_{h}^{(l)} = \bm{Q}^{(l)} \bm{W}_h^{Q(l)},
    \bm{K}_{h}^{(l)} = \bm{K}^{(l)} \bm{W}_h^{K(l)},
    \bm{V}_{h}^{(l)} = \bm{V}^{(l)} \bm{W}_h^{V(l)}
\end{equation}
where $\bm{Q}^{(l)} = \bm{K}^{(l)} = \bm{V}^{(l)} = \bm{H}^{(l-1)}$. In this paper $\bm{H}^{(l-1)}$ is the output of the encoder layer $(l-1)$ or the output of the embedding layer before the encoder layer 1. Matrices $\bm{Q}_{h}^{(l)}$, $\bm{K}_{h}^{(l)}$, and $\bm{V}_{h}^{(l)}$ can be further expressed as
\begin{equation}
\bm{Q}_{h}^{(l)} = {
\left( \begin{array}{c}
\bm{q}_0\\
\vdots\\
\bm{q}_M
\end{array}
\right )}, \,
\bm{K}_{h}^{(l)} = {
\left( \begin{array}{c}
\bm{k}_0\\
\vdots\\
\bm{k}_M
\end{array}
\right )}, \,
\bm{V}_{h}^{(l)} = {
\left( \begin{array}{c}
\bm{v}_0\\
\vdots\\
\bm{v}_M
\end{array}
\right )}
\end{equation}
where $\forall{\bm{q}, \bm{k}, \bm{v}} \in \mathbb{R}^{d_\text{v}}$. Then, we can obtain the scaled attention scores
\begin{align}
\frac{\bm{Q}_{h}^{(l)}\left(\bm{K}_{h}^{(l)}\right)^{T}}{\sqrt{d_{\text{v}}}} & = \frac{1}{\sqrt{d_{\text{v}}}}{
\left( \begin{array}{cccc}
\left(\bm{q}_0, \bm{k}_0\right) &\dots &\left(\bm{q}_0, \bm{k}_M\right)\\
\left(\bm{q}_1, \bm{k}_0\right)  &\dots &\left(\bm{q}_1, \bm{k}_M\right)\\
\dots & \left(\bm{q}_i, \bm{k}_j\right) &\dots\\
\left(\bm{q}_M, \bm{k}_0\right) &\dots &\left(\bm{q}_M, \bm{k}_M\right)\\
\end{array}
\right )} \nonumber \\ & = {
\left( \begin{array}{cccc}
u_{00} & \dots & u_{0M} \\
u_{10} &\dots & u_{1M}\\
\dots  & u_{ij} &\dots\\
u_{M0}  &\dots & u_{MM}\\
\end{array}
\right )}
\end{align}
where $(\bm{q}_i, \bm{k}_j), i,j \in \{0, \dots, M\}$ is the inner product of vectors, which measures the similarity of vector $\bm{q}_i$ and vector $\bm{k}_j$. The row-wise \textit{softmax} function is used on each element of the above scaled attention scores matrix, which is given by $\overline{u}_{ij} = e^{u_{ij}}/\sum_{j'=0}^{M}e^{u_{ij'}}$. Then, the output $\bm{Z}_h^{(l)} \in \mathbb{R}^{(M+1)\times d_{\text{v}}}$ of the $h$-th head is expressed as
\begin{align}
\bm{Z}_h^{(l)}  = {
\left( \begin{array}{c}
\sum_{j=0}^{M}\overline{u}_{0j}\bm{v}_{j}\\
\sum_{j=0}^{M}\overline{u}_{1j}\bm{v}_{j}\\
\vdots\\
\sum_{j=0}^{M}\overline{u}_{Mj}\bm{v}_{j}
\end{array}
\right ) = {
\left( \begin{array}{c}
\overline{\overline{\bm{u}}}_{0}\\
\overline{\overline{\bm{u}}}_{1}\\
\vdots\\
\overline{\overline{\bm{u}}}_{M}
\end{array}
\right )}}.
\end{align}

\begin{figure}[!t]
		\centering
		\includegraphics[width=3in]{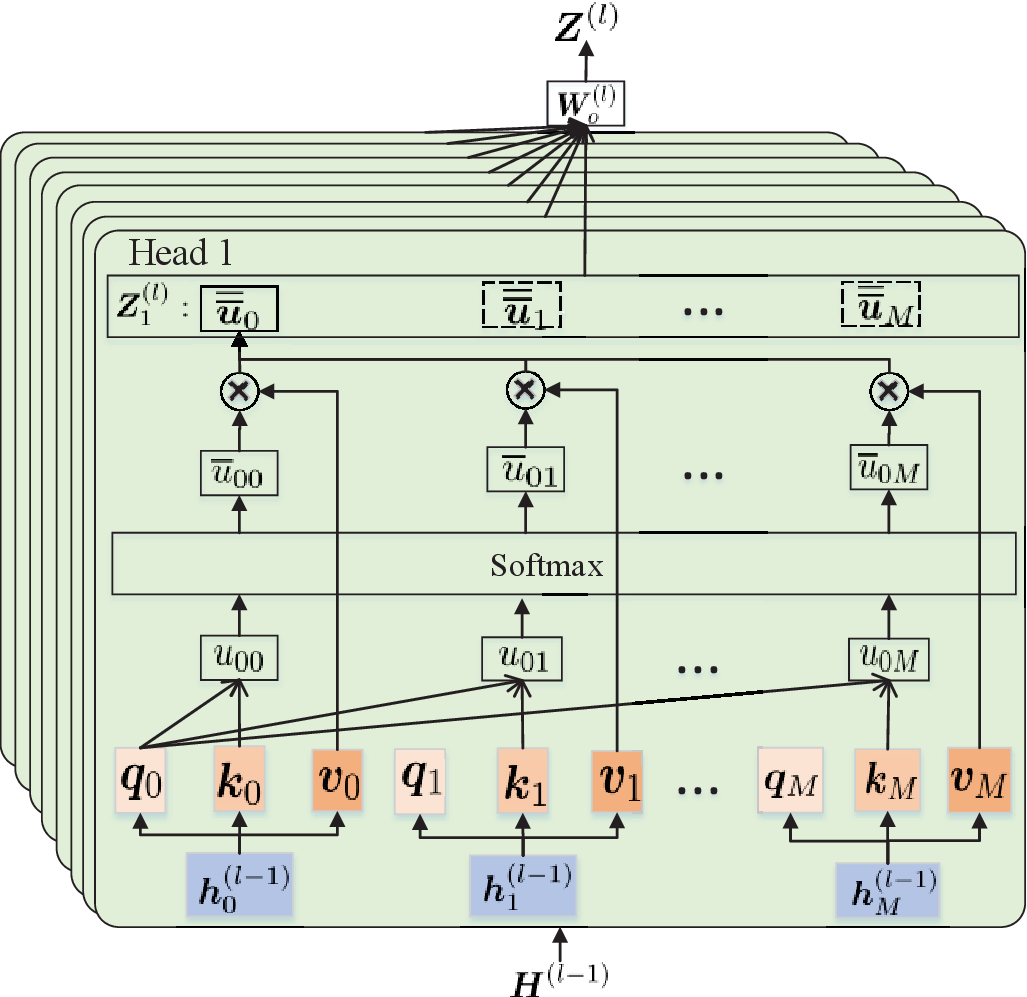}
		\caption{Multi-head self attention.}
		\label{mha}
	\end{figure}

Hence, we end up with 8 different outputs from 8 heads where each head could learn something different. These outputs are concatenated and multiplied by an additional learnable weight matrix $\bm{W}^{(l)}_{\rm o} \in \mathbb{R}^{8d_{\text{v}}\times d_{\text{em}}}$ to generate the final output of the multi-head attention layer, as follows:
\begin{equation}
   \label{outputofself}
    \bm{Z}^{(l)} = \left(\bm{Z}_1^{(l)},\dots, \bm{Z}_8^{(l)} \right) \bm{W}^{(l)}_{\rm o}, \, \bm{Z}^{(l)} \in \mathbb{R}^{(M+1) \times d_{\text{em}}}.
\end{equation}

To facilitate the understanding of multi-head attention layer, all operations from (\ref{singleattention}) to (\ref{outputofself}) are defined as a function $\text{MHA}(\cdot)$. Thus, $\bm{Z}^{(l)} = \text{MHA} \left(\bm{Q}^{(l)},  \bm{K}^{(l)}, \bm{V}^{(l)} \right)$.  Then, $\bm{Z}^{(l)}$ is added to the input of the multi-head attention in this encoder, which is a residual connection operation. Subsequently, the output of the residual connection is fed into a batch normalization, defined as a function BN($\cdot)$, and it is written as $\bm{Z}^{'(l)}  = \text{BN} \left(\bm{H}^{(l-1)} + \bm{Z}^{(l)}\right)$.
The use of the residual connection is to avoid the degradation problem of the network in training, while the layer normalization can improve the training speed and the stability of the networks.  The normalized residual output goes through a pointwise feed-forward network (defined as a function FFN($\cdot)$), which is a couple of linear layers with a ReLU activation in between. Then, the output of the pointwise feed-forward network is added to its input by a residual connection and further normalized to obtain the final output $\bm{H}^{(l)} \in \mathbb{R}^{(M+1) \times d_{\text{em}}}$ of the encoder layer $l$, which is given by $\bm{H}^{(l)} = \text{BN}\left(\bm{Z}^{'(l)} + \text{FFN} \left(\bm{Z}^{'(l)} \right)\right)$.
In each encoder layer, we perform the same computational process and finally output the final result of the encoder part at layer 6, $\bm{H}^{(6)} = \left(\bm{h}_0^{(6)}; \bm{h}_1^{(6)};\dots;\bm{h}_M^{(6)}\right) \in \mathbb{R}^{(M+1) \times d_{\text{em}}}$, which is the continuous representation with attention information of the input $\bm{H}^{({\rm in})}$. All of these operations will help the decoder network focus on the appropriate elements in the input during the decoding process.

\subsection{Decoder}
The decoding is autoregressive and generates the result one by one. The output of the decoder network can be represented as an ordered sequence of the input of the encoder. The decoder begins with the start point at decoding step $0$ since the trajectory of the UAV should start at the start point as well as end at this point. The output of each decoding step is based on the information from the encoder and the already-generated previous output in the decoder. Hence, the decoding process can be modelled using the probability chain rule:
\begin{align}
    \label{chainrule}
    P(&\bm{\pi}|\bm{H}^{({\rm in})}) \nonumber \\ & = \prod_{t=0}^{M+1}P(\pi(t)|\pi(0),\dots,\pi(t-1), \bm{H}^{({\rm in})}).
\end{align}
The decoding process aims at finding the optimal $\bm{\pi}$ to maximize $P(\bm{\pi}|\bm{H}^{({\rm in})})$.

The decoder network is composed of two identical decoder layers, and a single-head attention layer. Each decoder layer contains two multi-head attention sub-layers which employ a residual connection around them followed by layer normalization. These sub-layers have the same structure as the sub-layers in the encoder network but each of them has a different job. Since the output of the decoder network is related to the order, we need to inject some information about the positions into the input sequence of the decoder network. The locations are implicitly represented by the order of the data input to the decoder network. Hence, the input of the decoder network is the output of the encoder network combined with the positional encoding. Suppose the outputs of the decoder network at previous $t$ decoding steps are $\pi(0), \pi(1),\dots, \pi(t)$, the decoder wants to predict the output at $t+1$ step. Then, the input to the decoder network is expressed as $\widehat{\bm{H}}^{(0)}_{t+1} = \left(\widehat{\bm{h}}^{(0)}_{\pi(0)}; \widehat{\bm{h}}^{(0)}_{\pi(1)}; \dots; \widehat{\bm{h}}^{(0)}_{\pi(t)} \right)$. Each element in $\widehat{\bm{H}}^{(0)}_{t+1}$ can be calculated by $\widehat{\bm{h}}^{(0)}_{\pi(t)} = {\bm{h}}^{(6)}_{\pi(t)} + \text{PE}_{t}$,
where ${\bm{h}}^{(6)}_{\pi(t)} \in \mathbb{R}^{1 \times d_{\text{em}}}$ is one element in $\bm{H}^{(6)}$ which is decoded at the $t$-th step, $\text{PE}_{t} \in \mathbb{R}^{1 \times d_{\text{em}}}$ is the positional encoding based on the sinusoidal function, which is given by \cite{lin2021survey}

\begin{equation}
    \text{PE}_t(d_i) = \left\{ \begin{array}{ll}
     \sin{\left(\omega_{d_i} t\right)}, & \text{if $d_{i}$ is even} \\
     \cos{\left(\omega_{d_i} t\right)}, &  \text{if $d_{i}$ is odd}
      \end{array} \right.
\end{equation}
where $d_i$ is the dimension, $1 \leqslant d_i \leqslant d_{\text{em}}$, $\omega_{d_i}$ is the hand-crafted frequency for each dimension. The position encoding of each position successfully provides the position information to the decoder network. The input $\widehat{\bm{H}}^{(0)}_{t+1}$ gets fed into the first multi-head attention sub-layer of the first decoder layer and pass through the residual connection and layer normalization (denoted as a function LN($\cdot$)) to prepare the query for the next multi-head attention sub-layer, as follows
\begin{align}
    \widehat{\bm{Z}}^{(1)}_{t+1} &= \text{MHA}\left(\widehat{\bm{h}}^{(0)}_{\pi(t)}, \widehat{\bm{H}}^{(0)}_{t+1}, \widehat{\bm{H}}^{(0)}_{t+1}\right), \widehat{\bm{Z}}^{(1)}_{t+1} \in \mathbb{R}^{1 \times d_{\text{em}}}\\
    \widehat{\bm{Z}}^{'(1)}_{t+1} &= \text{LN} \left(\widehat{\bm{h}}^{(0)}_{\pi(t)} + \widehat{\bm{Z}}^{(1)}_{t+1}\right), \widehat{\bm{Z}}^{'(1)}_{t+1} \in \mathbb{R}^{1 \times d_{\text{em}}}
\end{align}
where $\widehat{\bm{h}}^{(0)}_{\pi(t)}$ is the query, $\widehat{\bm{H}}^{(0)}_{t+1}$ works as the key and the value matrices in the current multi-head attention sub-layer. The second multi-head attention sub-layer is used to match the encoder's input to the decoder's input to allow the decoder network to decide the next possible output among the non-visited elements. For this sub-layer, the encoder network's output $\bm{H}^{(6)}$ is the key and the value matrices, and $\widehat{\bm{Z}}^{'(1)}$ is the query matrix. The calculations are given by
\begin{align}
    \widehat{\bm{Z}}^{''(1)}_{t+1} & = \text{MHA} \left(\widehat{\bm{Z}}^{'(1)}_{t+1}, \bm{H}^{(6)}, \bm{H}^{(6)}\right), \widehat{\bm{Z}}^{''(1)}_{t+1} \in \mathbb{R}^{1 \times d_{\text{em}}}\\
    \widehat{\bm{H}}^{(1)}_{t+1} & = \text{LN} \left(\widehat{\bm{Z}}^{'(1)}_{t+1} + \widehat{\bm{Z}}^{''(1)}_{t+1}\right), \widehat{\bm{H}}_{t+1}^{(1)} \in \mathbb{R}^{1 \times d_{\text{em}}}.
\end{align}
Note that we add the mask of visited elements to the scaled attention scores in this sub-layer.  Then, $\widehat{\bm{H}}^{(1)}_{t+1}$ goes through the second decoder layer to get the output $\widehat{\bm{H}}^{(2)}_{t+1} \in \mathbb{R}^{1 \times d_{\text{em}}}$. In order for the decoder network to compute output probabilities $P(\pi(t+1)|\pi(0),\dots,\pi(t), \bm{H}^{(6)})$, $\widehat{\bm{H}}^{(2)}_{t+1}$ and the output $\bm{H}^{(6)}$ of the encoder network get fed into a single-head attention to get a distribution over the non-visited elements, which is given by \cite{bresson2021transformer}
\begin{equation}
   \bm{P}_{t+1} = \text{softmax}\left(\tanh{\left(\frac{\widehat{\bm{Q}}_{t+1} \widehat{\bm{K}}^{T}_{t+1}}{\sqrt{d_{\text{em}}}} \odot \mathcal{M}_{t+1}\right)} \right)
\end{equation}
where $\widehat{\bm{Q}}_{t+1} = \widehat{\bm{H}}^{(2)}_{t+1} \widehat{\bm{W}}_1,
    \widehat{\bm{K}}_{t+1} = \bm{H}^{(6)} \widehat{\bm{W}}_2$,
$\widehat{\bm{W}}_1 \in \mathbb{R}^{d_{\text{em}} \times d_{\text{em}}}$ and $\widehat{\bm{W}}_2 \in \mathbb{R}^{d_{\text{em}} \times d_{\text{em}}}$ are learnable weight matrices, $\mathcal{M}_{t+1}$ is the mask of the visited elements considered in this layer, $\odot$ is the Hadamard product, and $\bm{P}_{t+1}\in \mathbb{R}^{1 \times (M+1)}$ is the distribution over the non-visited elements, which is composed of probability scores. Then, the output that will be selected is sampled from the distribution with three decoding methods:

\subsubsection{Greedy} At each decoding step, this method greedily selects the element with the largest probability $P(\pi(t+1)|\pi(0),\dots,\pi(t), \bm{H}^{(6)})$.

\subsubsection{Random Sampling} This method randomly samples $W_{\text{sampling}}$ solutions, where each solution includes fully visiting order, and selects the solution with the highest probability as the final result.

\subsubsection{Beam Search} This method chooses the top $W_{\text{beam}}$ possible solutions that have the highest probability at each step, where $W_{\text{beam}}$ is the beam width. Those $W_{\text{beam}}$ solutions will move to the next time step, and the process repeats. Then, we can obtain a tree of solutions of each step and the $\bm{\pi}$ that has the highest overall probability is picked as the final result.

We assume that the index of the highest probability score in $\bm{P}_{t+1}$
is selected with the greedy decoding as the output $\pi{(t+1)}$ at step $t+1$. Thus $\pi(t+1)$ points to the element at the same position of the input sequence $\bm{H}^{({\rm in})}$ of the encoder network, which is represented as $\bm{h}_{\pi(t+1)}^{({\rm it})}$. Then, the decoder network takes the encoding information of $\bm{h}_{\pi(t+1)}^{({\rm it})}$ from $\bm{H}^{(6)}$, i.e., $\bm{h}_{\pi(t+1)}^{(6)}$, and adds it with its position encoding to the list of the decoder input to continue decoding for the next step. Finally, we can obtain a set of the visiting order, $\bm{\pi}$. As shown in the example in Fig.~\ref{transformer}, $\bm{H}^{({\rm in})} = \left(c_0;\left(G_1, b_1, N_1\right);\left(G_2, b_2, N_2\right); \left(G_3, b_3, N_3\right);\left(G_4, b_4, N_4\right)\right)$ is the input to the encoder network and the decoder network outputs the final visiting order $\bm{\pi} = \{\pi(0), \pi(1), \pi(2), \pi(3), \pi(4)\}$ to elements in $\bm{H}^{({\rm in})}$.


\subsection{Selection of Hovering Points}

Given the visiting order $\bm{\pi}$, we know the visiting order to all hovering points clusters and construct a graph containing all of them, as illustrated in Fig.~\ref{transformer}. Each layer of the graph is composed of one hovering points cluster. Then, we will calculate the path with the minimal total AoI starting from the start point (marked as $\pi(0)$ in the visiting order), going through each cluster $G_m$, and ending at the clone of the start point (marked as $\pi(M+1)$ in the visiting order). To guarantee that at most one hovering point is selected from each cluster, we assume that all edges between possible hovering points of consecutive clusters to be directed by $\bm{\pi}$. We use the weighted A* search algorithm \cite{ebendt2009weighted} to quickly find the hovering point from each cluster to build the path with the minimal cost (total AoI). We assume that the UAV currently reaches the point $s{'}$ and will decide the next point to be expanded by the following cost function
\begin{equation}
    f(s) = g(s) + \omega h(s)
\end{equation}
where $s$ is any neighbor point of $s{'}$, $g(s)$ is the total movement cost on the path from the start point $c_0$ to $s$, $h(s)$ is the heuristic function to estimate cost from $s$ to the end point $c_0{'}$, and $\omega > 1$ is a constant factor. The neighbor point with a minimal $f(s)$ value is expanded. The pseudocode is described in Algorithm \ref{alg1}. We use COST and FRONTIER to keep track of $g(s)$ and the expanding process, respectively. Each point that has been reached keeps a pointer to its parent in CAME\_FROM so that we can know where it came from. With CAME\_FROM, we can construct a path having the minimal AoI from the start point to the end point, as illustrated by the solid red line with arrow in the example in Fig.~\ref{transformer}.

\begin{algorithm}[t!]
	\caption{Pseudocode for weighted A* search algorithm to find hovering points}
	\label{alg1}
	\begin{algorithmic}[1]
		\renewcommand{\algorithmicrequire}{\textbf{Input:}}
		\REQUIRE created graph\\
	     \STATE FRONTIER = PriorityQueue()\\
	     \STATE FRONTIER.put($c_0$, 0)\\
	     \STATE CAME\_FROM = $\left[\,\right]$ \\
	     \STATE COST = $\left[\,\right]$ \\
	     \STATE CAME\_FROM$\left[c_0\right]$ = None \\
	     \STATE COST$\left[c_0\right]$  = 0 \\
	     \WHILE{FRONTIER is not empty}
	           \STATE current point $s{'}$ =  FRONTIER.get() \\
	            \IF{$s{'} = c_0{'}$}
	               \STATE break
	            \ENDIF
	            \FOR{each neighbor $s$ of $s{'}$}
	               \STATE $g(s) = \text{COST}[ s{'}] +$ the total AoI from $s{'}$ to $s$
	               \IF{$s$ not in COST or $g(s) < \text{COST}[s]$}
	                   \STATE COST$[s]$ = g(s)
	                   \STATE $f(s) = g(s) + \omega h(s)$
	                   \STATE FRONTIER.put($s$, $f(s)$)
	                   \STATE CAME\_FROM$[s]$ = $s{'}$
	               \ENDIF
	            \ENDFOR
	     \ENDWHILE
	     \STATE calculate $\overline{A}$ according to CAME\_FROM
	\end{algorithmic}
    \end{algorithm}

\begin{algorithm}[!t]
	\caption{Training TWA* by REINFORCE with rollout baseline}
	\label{trainingprocess}
	\begin{algorithmic}[1]
		\renewcommand{\algorithmicrequire}{\textbf{Input:}}
	     \REQUIRE Epochs $E_{\text{epochs}}$, training steps $S$, batch size $B_{\text{size}}$
	    \STATE Initialize parameters $\vartheta$, $\vartheta^{({\rm BL})} \gets \vartheta$\\
	    \FOR{epoch = 1 to $E_{\text{epochs}}$}
	    \FOR{step = 1 to $S$}
	     \STATE $\bm{H}^{({\rm in})}_i \gets$ generate instances() $\forall{i \in \{1,\dots, B_{\text{size}}\}}$\\
	     \STATE $\bm{\pi}_i \gets$ Sampling solution$P_{\vartheta}\left(\cdot|\bm{H}^{({\rm in})}_i\right)$\\
	     \STATE $\bm{\pi}_i^{({\rm BL})} \gets$ Greedy solution$P_{\vartheta^{({\rm BL})}}\left(\cdot|\bm{H}^{({\rm in})}_i\right)$\\
	     \STATE $\overline{A}_i \gets \text{weighted A*} \left(\bm{\pi}_i \right)$\\
	     \STATE $\overline{A}^{({\rm BL})}_i \gets \text{weighted A*} \left(\bm{\pi}^{({\rm BL})}_i \right)$
	     \STATE $\nabla_{\vartheta} J \gets \sum_{i=1}^{B_{\text{size}}}\left(\overline{A}_i-\overline{A}^{({\rm BL})}_i\right) \nabla_{\vartheta}\log P_{\vartheta}\left( \bm{\pi}_i | \bm{H}^{({\rm in})}_i\right)$\\
	     \STATE $\vartheta \gets \text{Adam}\left(\vartheta, \nabla_{\vartheta} J\right)$
	    \ENDFOR
	    \IF{$t$-test$\left(P_{\vartheta}(\cdot), P_{\vartheta^{({\rm BL})}}(\cdot)\right) < 5\%$}
	         \STATE $\vartheta^{({\rm BL})} \gets \vartheta$
	    \ENDIF
		\ENDFOR
	\end{algorithmic}
    \end{algorithm}

\subsection{Computational Complexity Analysis}
In the encoder network, each encoder layer is the standard transformer encoder with quadratic computational complexity $O((M + 1)^2 d_{\text{em}})$ \cite{vaswani2017attention}. Since the number of layers is constant, the computational complexity of the encoder network is still $O((M + 1)^2 d_{\text{em}})$. In the decoder network, although each decoder layer contains two multi-head attention sub-layers, its computational complexity is still estimated to be quadratic $O((M + 1)^2 d_{\text{em}})$ \cite{vaswani2017attention}. Likewise, the number of decoder layers does not affect the computational complexity of the decoder network. In addition, the computational complexity of the single-head attention used in the final step of the decoder network is also quadratic $O((M + 1)^2 d_{\text{em}})$ \cite{vaswani2017attention}. Hence, the employed transformer model has the computational complexity $O((M + 1)^2 d_{\text{em}})$,  which is quadratic in the length of the input sequence. Different data structures used to implement the weighted A* algorithm, and hence, affect its computational complexity. We use the min heap to implement the weighted A* algorithm. We assume that at most $ML_{\rm sub}^2$ points (the total number of points in the search graph) are visited, and the min heap uses $O(\log (ML_{\rm sub}^2))$ computational complexity to extract a point each time \cite{RAMALINGAM1996233}. The weighted A* algorithm's computational complexity is estimated to be $O(ML_{\rm sub}^2\log (ML_{\rm sub}^2)$. Hence, the computational complexity of the proposed algorithm is $O((M + 1)^2 d_{\text{em}}) + O(ML_{\rm sub}^2\log (ML_{\rm sub}^2)$.



\subsection{Training}

To enable the transformer model to produce the optimal $\bm{\pi}$, we use the well-known policy gradient approaches to train it. The transformer model is parameterized by $\vartheta$, which includes all trainable variables in the encoder and the decoder networks. We regard the UAV as an agent to learn a good policy $\bm{\pi}$ to maximize long-term rewards by iteratively interacting with the environment to optimize parameter $\vartheta$. At each step, the agent in a given state chooses an action by its decision policy, which actually is the mapping from states to actions.

\subsubsection{State} The state consists of the environment encoded by the encoder network and the visited clusters before the current step in the decoder, which are $\bm{H}^{(6)}$ and $\widehat{\bm{H}}^{(0)}_{t+1}$ in the transformer, respectively.

\subsubsection{Action} At each step, the agent makes an action $\pi(t)$ based on its state, which can be seen as the processes of the right-hand side of (\ref{chainrule}). Thus, we view all operations in the decoder network as the action.

\subsubsection{Reward} The negative of the total AoI $\overline{A}$ in (\ref{averageaoi}) is used as the reward.

Our objective for training is given by

\begin{equation}
\label{trainingobjective}
    J\left(\vartheta| \bm{H}^{({\rm in})}\right) = \mathbb{E}_{\bm{\pi}\sim P_{\vartheta} \left(\cdot|\bm{H}^{({\rm in})}\right)} \left( \overline{A}\right).
\end{equation}
The gradient of (\ref{trainingobjective}) is calculated using the REINFORCE algorithm \cite{williams1992simple} with the greedy rollout baseline $\overline{A}^{({\rm BL})}$ \cite{kool2018attention}

\begin{align}
	\label{gradient}
	    \nabla_{\vartheta} & J\left(\vartheta | \bm{H}^{({\rm in})}\right) \nonumber \\ 
	    & = \mathbb{E}_{\bm{\pi}\sim P_{\vartheta}\left(\cdot|\bm{H}^{({\rm in})}\right)  }\left[\left(\overline{A} - \overline{A}^{({\rm BL})}\right) \nabla_{\vartheta}\log P_{\vartheta}\left( \bm{\pi} | \bm{H}^{({\rm in})}\right) \right]
\end{align}
where $\overline{A}$ is the cost of a solution that is obtained from the current training transformer model by sampling decoding. We set the greedy policy as the baseline policy in our model, and hence, $\overline{A}^{({\rm BL})}$ is the cost of a solution of the deterministic greedy decoding, which is used to eliminate  variance during training. By doing so, the transformer model is trained to improve over its (greedy) self. The training process is summarized in Algorithm \ref{trainingprocess}. In each training step, new instances are generated first (line 4). Then, the transformer model uses sampling decoding and greedy decoding to produce $\bm{\pi}_i$ and $\bm{\pi}_i^{({\rm BL})}$ (lines 5 and 6), respectively. The total AoIs are further obtained from the weighted A* (lines 7 and 8). The gradient in (\ref{gradient}) is approximated with Monte Carlo sampling in a batch size $B_{\text{size}}$ (line 9). The model parameter $\vartheta$ is updated using the Adam optimizer (line 10). We compare the current policy with the greedy baseline policy and update the parameter $\vartheta^{({\rm BL})}$ only if the improvement is significant according to a paired $t$-test $(5\%)$ \cite{kool2018attention}.

As pointed out earlier, Ptr-A* proposed in our previous work \cite{9507262} can also be used to solve the formulated GTSP in this work. Here, we give comparisons between TWA* and Ptr-A* in detail. First, they have different structures. In Ptr-A*, the encoder network consists of the LSTM networks, and the decoder network is composed of the LSTM networks and the attention mechanism. The LSTM networks have the form of a chain of repeating modules of a neural network. The key to the LSTM networks is the cell state, which is the hidden state. In theory, the hidden state can carry relevant information throughout the processing of the sequence. Since the LSTM networks process the elements of the input sequence one by one, the hidden state of each element of the input sequence is calculated by the current element and the previous hidden state. The final hidden state of the encoder network is fed into the LSTM networks of the decoder network. Then, the attention mechanism uses the hidden state of the decoder network to generate the output sequence. In TWA*, the encoder network includes six identical encoder layers in which each encoder layer is mainly composed of one multi-head self-attention sub-layer and one point-wise feed-forward network. The decoder network of TWA* consists of two identical decoder layers and one single-head attention layer. The encoder network is used to map the input $\bm{H}^{(0)} = \left(\bm{h}^{(0)}_0; \bm{h}^{(0)}_1; \dots; \bm{h}^{(0)}_M \right)$ to a sequence of continuous representations. The decoder network receives the output of the encoder together with the decoder output at the previous time step to generate the output sequence. Multi-head self-attention is an attention mechanism relating different positions of a single sequence in order to compute a representation of the sequence. Multi-head self-attention helps TWA* to look at all elements in the input sequence for clues that can help lead to a better encoding. Unlike hidden states used in Ptr-A*, TWA* does not rely on past hidden states to capture dependencies with previous elements in the sequence. As a result, TWA* does not suffer from long dependency issues, which are very common in recurrent-based networks, such as RNNs and LSTMs, and hence does not lost past information.

Second, TWA* can process sequences in parallel, which is faster than Ptr-A*. In Ptr-A*, the elements of a sequence must be processed one by one and each element's hidden state is assumed to be dependent only on the previously hidden state. Hence, Ptr-A*'s recurrent structure makes it hard to use parallel computing to process sentences and this means that it is very slow in training and inference. All elements in a sequence are processed in TWA* as a whole rather than one by one. Because of the use of multi-head self-attention that is designed in parallel, TWA* has the ability of parallel computation.

Third, the training method used in this work is different from the training method in \cite{9507262}. In \cite{9507262}, in order to obtain the optimal parameter of Ptr-A*, we use the actor-critic architecture to train Ptr-A* where a second critic network must be trained. In this work, we use the greedy rollout baseline where the TWA* model is trained to improve over its (greedy) self. This training method can avoid all the inherent training difficulties associated with the actor-critic architecture.

\section{Numerical Results}\label{SecIV}

We conduct extensive experiments to investigate the performance of the proposed TWA* algorithm in solving the problem of trajectory planning to minimize the total AoI for the UAV-IoT network. The proposed model is implemented by Pytorch 1.7 and Python 3.8 and trained on a machine with 1 NVIDIA RTX 2080Ti GPU.

\subsection{Test Settings}

\subsubsection{Decoding Strategies}
As we mentioned in Section \ref{SecIII}-D, the random sampling decoding and the greedy decoding are employed for training the model. At inference, we evaluate performance of all three decoding methods on test instances and they are marked as TWA*--greedy, TWA*--sampling ($W_{\text{sampling}} = 5120$), and TWA*--beam search ($W_{\text{beam}}=100$).

\subsubsection{Comparison Algorithms}

To evaluate the effectiveness of the proposed model with different decoding methods, we compare it with the genetic algorithm \cite{yang2008solving}, the simulated annealing (SA) algorithm \cite{zhan2016list}, and Ptr-A* with the sampling strategy \cite{9507262}. Common parameters are selected for the genetic algorithm: The population size is the number of all possible hovering points in one instance, the maximal iteration is 10000, crossover is 0.1, and mutation probability is 0.8. The parameters of SA for the initial temperature, cooling coefficient, and maximal iteration are taken as 100, 0.99, and 1000, respectively.

\subsubsection{Data Generation}
We assume there is a probability distribution over a family of problems. During training, problem instances are generated according to this distribution, and any test examples are also produced from the same distribution at inference. For any problem instances, all CHs $\{b_1,\dots,b_M\}$ are randomly sampled from the distribution $\bm{U} = {\tt torch.FloatTensor(1, 2).uniform(0, 3000)}$. With the SNR threshold $\gamma_{\text{th}}$ and environment parameters, we can calculate the hovering disk $O_m$ for each CH $b_m$, and each cluster of candidate hovering points $G_m$ is sampled from $O_m$. The number of nodes $N_m$ in each ground cluster is randomly chosen from $\{5,10,15,20,25,30\}$. Hence, any of the problem instances is obtained as $\bm{H}^{({\rm in})} = \left(c_0; \left(G_1, b_1, N_1\right);\dots; \left(G_m, b_m, N_m\right);\dots;\left(G_M, b_M, N_M\right) \right)$.

\begin{table}[t!]
	\centering
	\caption{Simulation parameters}
	\label{table1}
	\footnotesize
	 \begin{tabular}{p{1.3cm}<{\centering}|p{2.2cm}<{\centering}||p{1.3cm}<{\centering}|p{2.3cm}<{\centering}}
		\hline{}
		\textbf{Parameter} & \textbf{Value} & \textbf{Parameter} & \textbf{Value}\\
		\hline\hline
		$H$ & 100 m  & $\beta$ & 12.08\\
		\hline
		$\widetilde{\beta}$ & 0.11 & $P_{\text{CH}}$ & 0.1 W\\
		\hline
		$\gamma_{\text{th}}$ & 20 dB (default) & $\xi_{\text{LoS}}$ & 1 dB\\
		\hline
		$\xi_{\text{NLoS}}$  & 20 dB &  $\sigma^2$ & $-110$ dBm \\
        \hline
        $v_{\text{UAV}}$ & 15 m/s & $f_c$ & 2 GHz\\
        \hline
        $L_{\text{data}}$ & 5 Mb & $B_{\text{width}}$ & 1 MHz \\
        \hline
        $P_{\text{com}}$ & 0.1 W & $L_{\text{sub}}$ & 5\\
        \hline
        $P_0$ & 99.66 W & $P_1$ & 120.16 W\\
        \hline
        $U_{\text{tip}}$ & 120 m/s & $v_0$  & 0.002 m/s\\
        \hline
        $d_0$ & 0.48 & $\rho$ & 1.225 $\text{kg/m}^3$\\
        \hline
        $\tau$ & 0.1 s & $\delta$ & 0.5 \\
        \hline
        $s_0$ & 0.0001 & & \\
        \hline
		\end{tabular}
\end{table}

\subsubsection{Environment Parameters and Hyperparameters}

We consider a ground network with a size of 3 km $\times$ 3 km, and the start position of the UAV is located at $(0\,\text{m}, 0\,\text{m}, H\,\text{m})$. Environment parameters are listed in Table \ref{table1}. The embedding dimension $d_{\text{em}}$ is equal to 512 and $d_{\text{v}}$ is equal to 64. We train the proposed model using the Adam optimizer with a learning rate of 0.0001 on $E_{\text{epochs}} = 200$ epochs, where each epoch includes $S=1000$ training steps. At each training step, the batch size $B_{\text{size}}$ is equal to 512, which means there are 512 instances in each batch.  In each instance, we set $M=10$.

\subsection{Analysis of the Results}
    
	\begin{figure}[!t]
      \centering
      \subfigure[Comparison of the total AoI when $M$ varies.]{\includegraphics[width=3.0in]{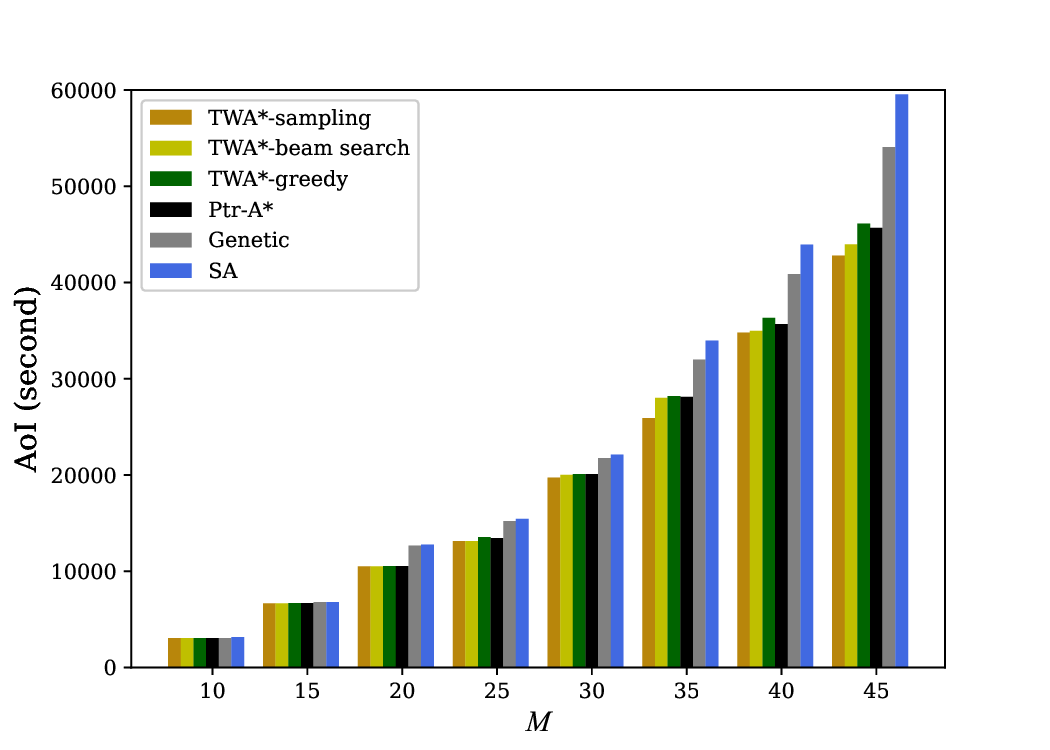}}
	  \subfigure[Comparison of the oldest packet's AoI when $M$ varies.]{\includegraphics[width=3.0in]{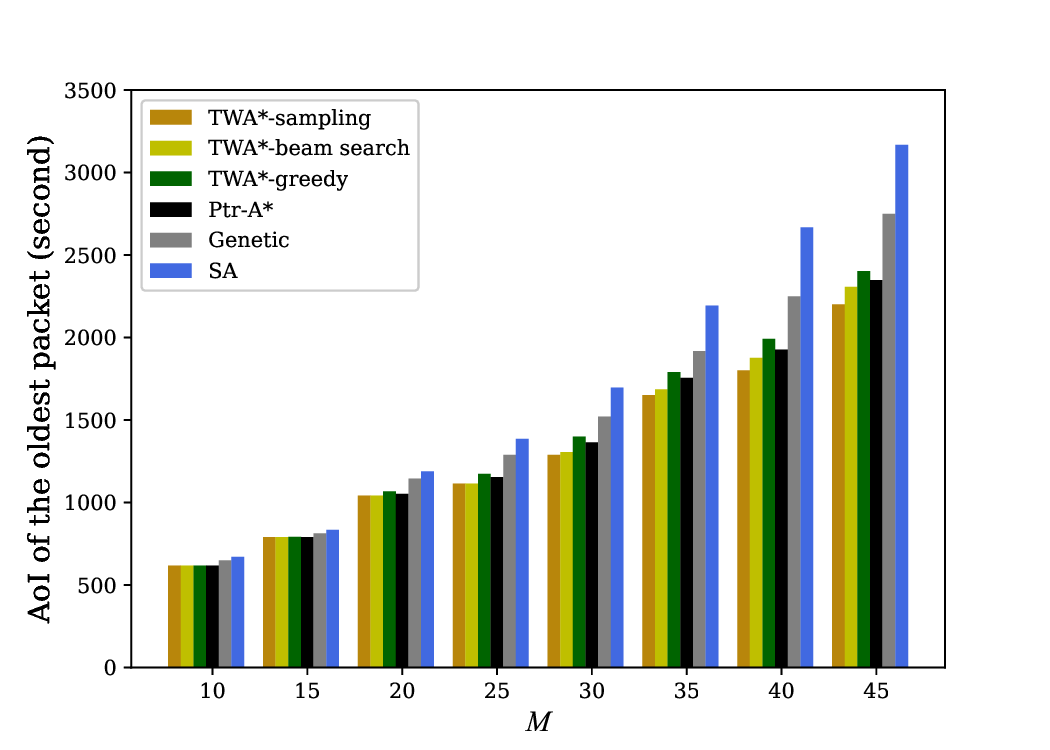}}
     \caption{Comparison when $M$ varies.}
     \label{aoicomparsion}
   \end{figure}

	\begin{figure}[!t]
		\centering
		\includegraphics[width=3.0in]{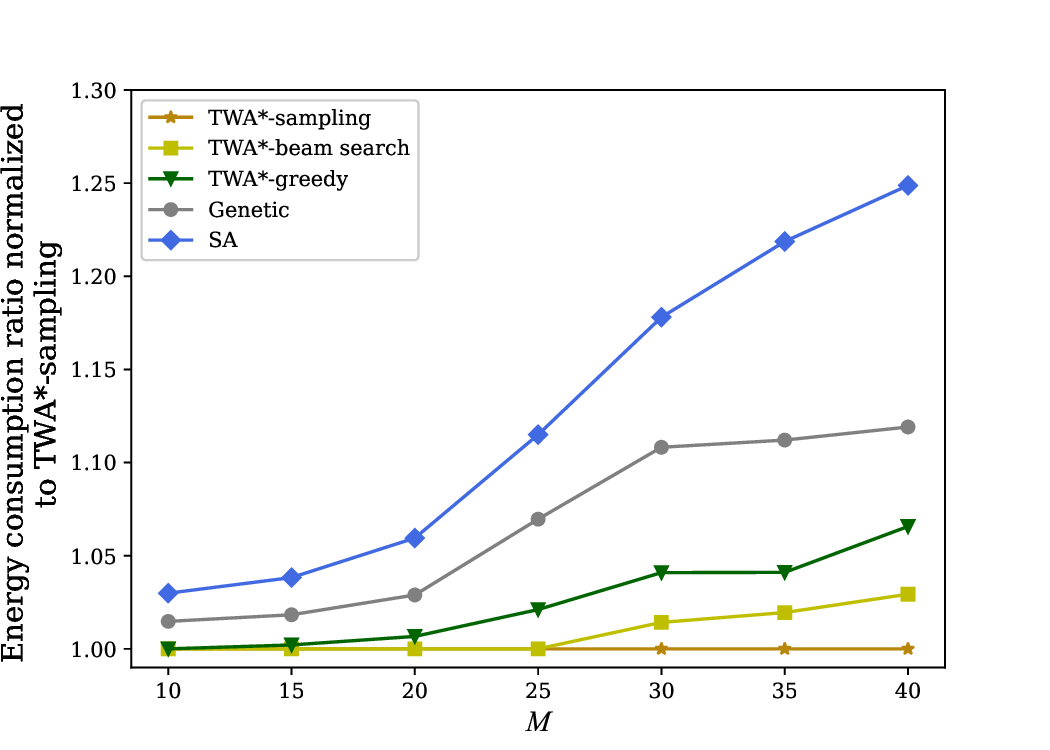}
		\caption{Comparison of energy consumption when $M$ varies.}
		\label{energyratio}
	\end{figure}
	
We first compare the total AoI between our proposed algorithm against the genetic and SA algorithms on the trained model  when the value of $M$ varies. Although the model is trained on 10-clusters IoT networks ($M=10$), it still shows good performance on IoT networks with different sizes, like 20-clusters, 30-clusters, etc., as can be seen in Fig.~\ref{aoicomparsion} (a). Specifically, the TWA*-sampling algorithm always obtains the minimal total AoI when compared with other two decoding methods, as well as the three other algorithms under comparison. The TWA*-beam search and TWA*-greedy algorithms exhibit an obviously superior performance than the genetic and SA algorithms in reducing the total AoI. The above observations indicate that the proposed algorithm with the three different decoding methods achieves an excellent generalization ability with respect to the size of the IoT network used for training. When $M=10$, TWA*-sampling, TWA*-beam search, TWA*-greedy, Ptr-A*, and the genetic algorithms obtain almost the same total AoI; however, the SA algorithm has a higher total AoI when compared to our proposed algorithm with all three decoding strategies. As the value of $M$ increases, the performance gap increases gradually between our proposed algorithm and comparison algorithms. For instance, when $M = 25$, the total AoI values of the TWA*-sampling, TWA*-beam search, TWA*-greedy,  Ptr-A*, genetic, and SA algorithms are 13134, 13134, 13546, 13431, 15205, and 15452 seconds, respectively. Compared to Ptr-A*, TWA*-sampling has a performance gain of $2.2\%$. As $M$ increases to 45, the total AoI values obtained by the TWA*-sampling, TWA*-beam search, and TWA*-greedy algorithms are 42803, 43971, and 46118 seconds, respectively. Compared to Ptr-A* with a AoI value of 45663 seconds, TWA*-sampling has a performance gain of $6.3\%$. However, the total AoI values of the genetic and SA algorithms are 54061 and 59537 seconds, respectively, which are obviously inferior than what obtained by our proposed algorithm. In summary, our proposed algorithm using any of the three decoding methods can obtain better total AoI results than both the genetic and SA algorithms. In addition, TWA*-sampling always obtains better AoI values than Ptr-A* with the sampling strategy. This comparison result is consistent with the conclusions in \cite{kool2018attention} and \cite{bresson2021transformer} that the transformer-based technique outperforms the pointer network-based technique.

Next, we compare the AoI of the oldest packet which is from the node $b_{\pi(1)}^{(1)}$ that samples data first in the whole IoT network, among different algorithms. As can be seen in Fig.~\ref{aoicomparsion} (b), our proposed algorithm also exhibits a good performance in reducing the AoI of the oldest packet when compared with the genetic and SA algorithms. Furthermore, the TWA*-sampling algorithm obtains the best results among the three decoding methods and Ptr-A*.

 \begin{table*}[t!]
	\centering
	\caption{Comparison of running time (second).}
	\label{timesmall}	 \begin{tabular}{p{0.5cm}<{\centering}|p{2.15cm}<{\centering}|p{2.55cm}<{\centering}|p{2.0cm}<{\centering}|p{2.0cm}<{\centering}|p{1cm}<{\centering}|p{1cm}<{\centering}}
		\hline
	    & \multicolumn{5}{c}{Algorithm} \\
		\hline
	    $M$ & TWA*-sampling & TWA*-beam search & TWA*-greedy &  {Ptr-A*} & Genetic & SA\\
		\hline\hline
	    10 & 1.9693 & 2.0653 & 1.9556 & {11.1556} & 47.57 & 5.5345\\
		\hline
		15 & 2.1412 & 2.2861 & 2.1055 & {19.3212} & 95.46 & 6.4262\\
		\hline
		20 & 2.3392 & 2.4900 & 2.3037 & {27.9023} & 163.41 & 6.8623\\
		\hline
		25 & 2.6006 & 2.8087 & 2.5778  & {36.9560} & 261.69 &  7.4619\\
	    \hline
	    30 & 2.8700 & 3.0876 & 2.8300 & {43.5498} & 392.97 & 8.3378\\
	    \hline
	    35 & 3.2018 & 3.4531 & 3.1536 & {52.4981} & 562.25 & 9.2576\\
	    \hline
	    40 & 3.8059 & 3.7506 & 3.5583 & {61.6301} & 749.16 & 9.6988\\
	    \hline
	    {45} & {4.5112} & {4.5190} & {3.8995} & {75.8817} & {991.85} & {10.2019}\\
	    \hline
		\end{tabular}
   \end{table*}

Given that the proposed algorithm can find the best UAV trajectory with the minimal AoI in the UAV-IoT network among all the algorithms under comparison, it is of interest to further investigate the effective energy consumption of the UAV. It can be seen from (\ref{consumption_hovering}) and (\ref{consumption_flying}) that the energy consumption of the UAV is related to its flying time and hovering time. The effective energy consumption is defined as the energy consumption of the UAV from the first visited hovering point to the end point, i.e., in completing its data collection task. Fig.~\ref{energyratio} compares the effective energy consumptions for all the algorithms for different values of $M$. In particular, plotted in the figure are the average ratios of the effective energy consumptions by different algorithms with over that of the TWA*-sampling algorithm. As can be seen, our proposed algorithm with any of the three decoding methods has a better performance when compared to the genetic and SA algorithms, whereas the TWA*-sampling algorithm obtains the best result. The results in Fig.~\ref{energyratio} are in line with expectations because with our proposed algorithm, the UAV spends less time to gather data than with the other two algorithms, which helps to reduce the effective energy consumption of the UAV.

	
	
    \begin{figure}[!t]
      \centering
      \subfigure[Comparison of the total AoI for different values of $\gamma_{\text{th}}$.]{\includegraphics[width=3.0in]{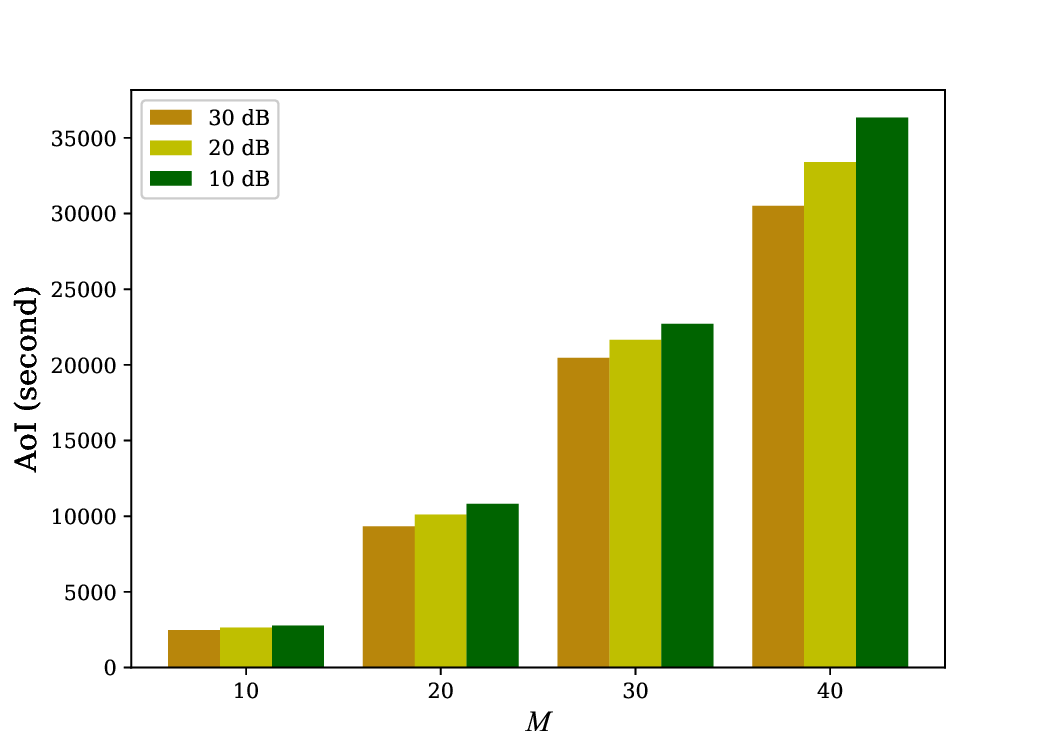}}
	  \subfigure[Percentages of the total flying time and the total hovering time for different values of $\gamma_{\text{th}}$.]{\includegraphics[width=3.0in]{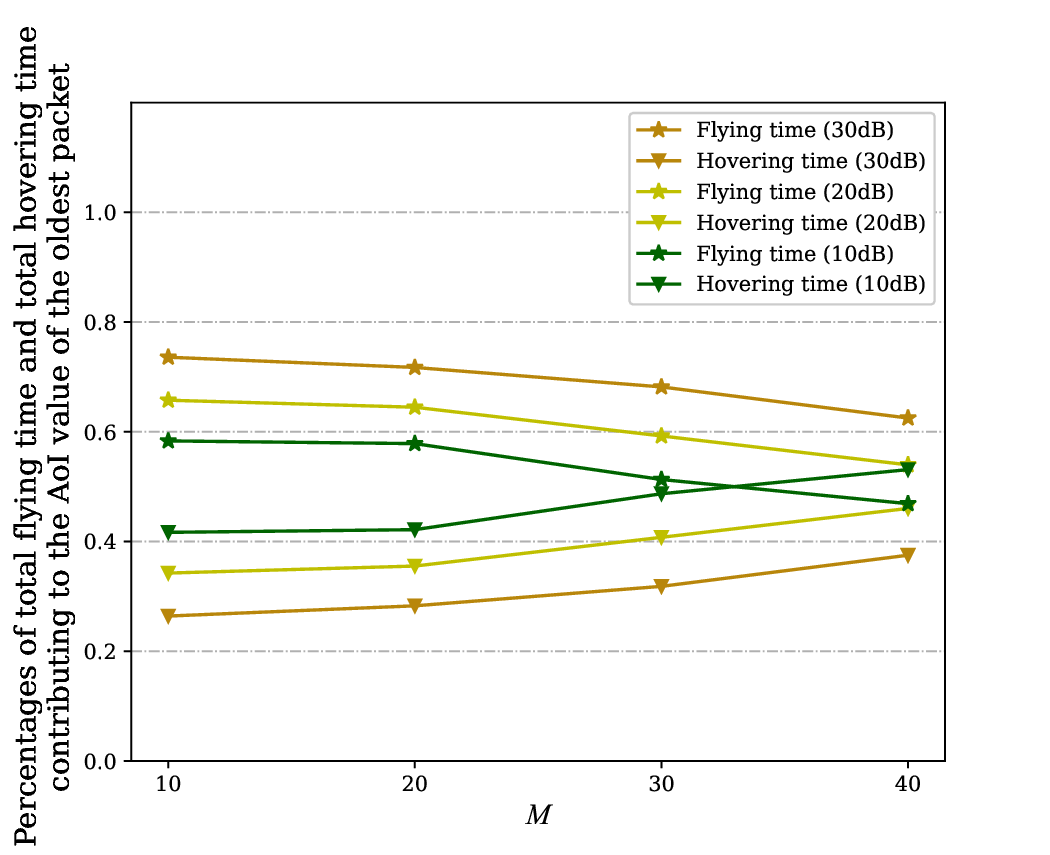}}
     \caption{Comparison for different values of $\gamma_{\text{th}}$.}
     \label{comparisongama}
   \end{figure}

Table \ref{timesmall} compares the running time at inference. As $M$ increases, the running time of all the algorithms increases, which is well expected. We can see that among all the algorithms and for all the values of $M$, the running time of the TWA*-greedy algorithm is always shortest. Although TWA*-sampling obtains the best performance in reducing the total AoI as well as the AoI of the oldest packet (as can be seen from Fig.~\ref{aoicomparsion} (a) and Fig.~\ref{aoicomparsion} (b)), it has a longer running time than TWA*-greedy, which is reasonable. Similarly, TWA*-beam search takes slightly more time than TWA*-greedy because it needs more computational time to search for a better solution than TWA*-greedy. We can observe that the genetic algorithm takes the longest time among all the algorithms and its running time significantly increases as $M$ increases. The running time of SA is acceptable in comparison with the genetic algorithm. Overall, the computational performance of our proposed model with all three decoding methods is significantly better than the SA and genetic algorithms. The running time of Ptr-A* is much greater than that of TWA*-sampling, TWA*-beam search, and TWA*-greedy. For example, when $M = 20$, the running time of the Ptr-A* is 11.9 times that of TWA*-sampling. When $M$ increases to 45, the running time of the Ptr-A* is 16.8 times that of TWA*-sampling. This is because transformer-based techniques can process a sequence in parallel. However, with Ptr-A*, the elements of a sequence must be processed one by one. Hence, our proposed TWA* with three decoding methods is faster than Ptr-A*.

In order to provide insights about the effect of $\gamma_{\text{th}}$ on the total AoI, we set the same number of devices, namely $N_m=20$, in each ground cluster and evaluate in Fig.~\ref{comparisongama} (a) the performance of TWA*-sampling for different values of $\gamma_{\text{th}}$. According to \emph{Lemma 1} and (\ref{diskregion}), the smaller the value of $\gamma_{\text{th}}$ is, the larger the area of each hovering disk $O_m$ will be. This will affect the positions of hovering points and thus the total AoI. As we can see in Fig.~\ref{comparisongama} (a), for any given number of ground clusters, the total AoI gradually increases as the value of $\gamma_{\text{th}}$ decreases. For example, when $M=30$, the values of total AoI in 10 dB, 20 dB, and 30 dB are 22707, 21655, and 20457 seconds, respectively. We can also observe that the total AoI gap among three values of $\gamma_{\text{th}}$ increases as $\gamma_{\text{th}}$ becomes higher.

Next, we compare the total flying time and the total hovering time of the UAV that make up of the AoI value $A_{\pi(1)}^{(1)}$ of the oldest packet in networks with different number of clusters. Specifically these total flying time and total hovering time of the UAV are calculated as $\sum^{M}_{t=1}T^{({\rm fly})}_{(c_{\pi(t)}, c_{\pi(t+1)})}$ and $\sum^{M}_{t=1} T^{({\rm hov})}_{c_{\pi(t)}}$, respectively. In each network, we also compare these portions of time when $\gamma_{\text{th}}$ varies. Note that, the AoI values of the oldest packet are different for different $\gamma_{\text{th}}$ values in a network. The percentages of the total hovering and the total flying time that contribute to the AoI value of the oldest packet are plotted in Fig.~\ref{comparisongama} (b). When $M=10$, the total flying time is always higher than the total hovering time for any thresholds $\gamma_{\text{th}}$. In addition, as $\gamma_{\text{th}}$ increases, the flying time portion increases. This is because the selected hovering point may be closer to the center of each hovering disk if the value of $\gamma_{\text{th}}$ is large, which will cause the flight distance to be longer and thus increases the total flying time. When $M$ increases, the UAV needs more time to collect data, and we can see that the hovering time portion slowly increases as expected.

    \begin{figure}[!t]
      \centering
      \subfigure[Comparison of the total AoI when $N$ varies]{\includegraphics[width=3.0in]{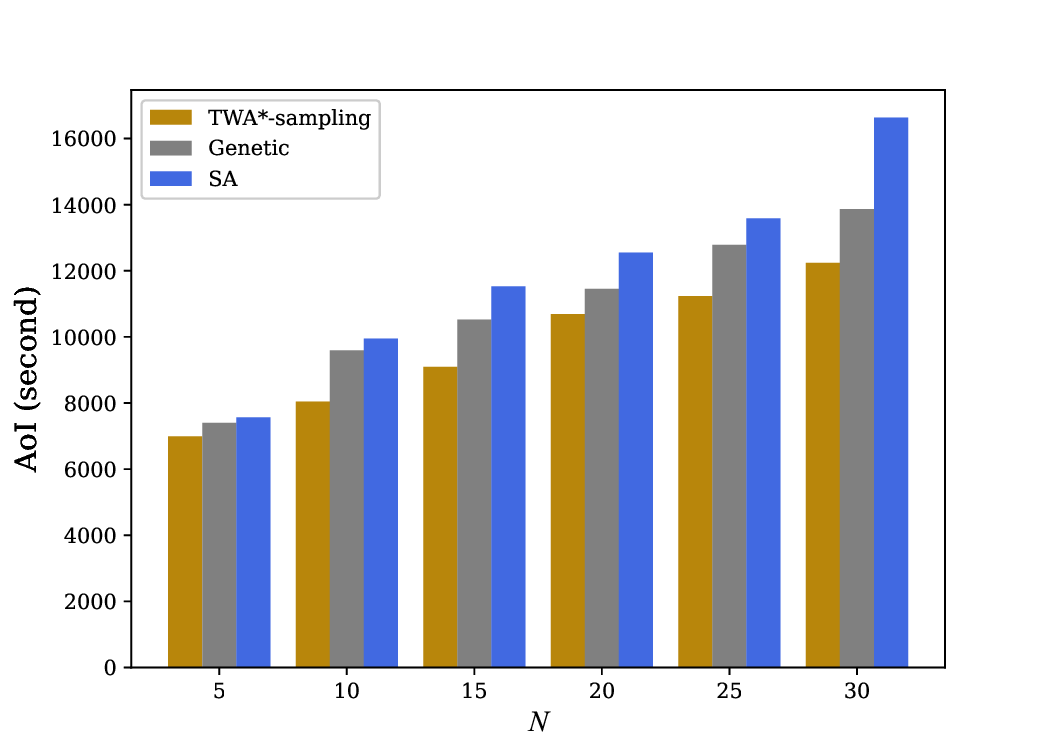}}
	  \subfigure[Percentages of flying time and hovering time when $N$ varies.]{\includegraphics[width=3.0in]{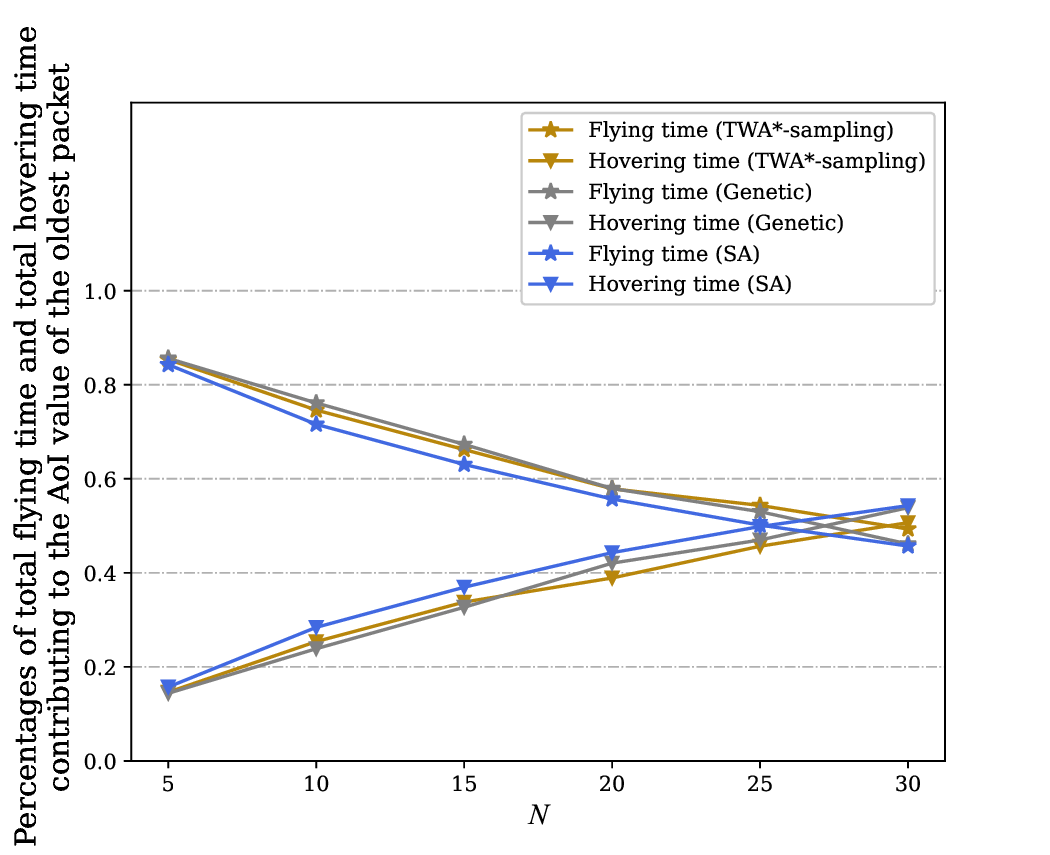}}
     \caption{Comparison when $N$ varies.}
     \label{comparison_n}
   \end{figure}	

Fig.~\ref{comparison_n} (a) compares the total AoI for different algorithms and for different numbers of devices in each cluster. We test the trained model on a 20-clusters instance with the same number of devices in each cluster. It can be seen that the proposed algorithm with the sampling decoding method always obtains the minimal values when compared to the genetic and SA algorithms. This clearly shows that our proposed algorithm can find a better trajectory in reducing the total AoI. As $N$ increases, there is a large performance gap between TWA*-sampling and the two algorithms under comparison.

Fig.~\ref{comparison_n} (b) plots the percentages of the total flying time and the total hovering time that make up of the AoI value of the oldest packet in the 20-clusters network when $N$ varies. For all the algorithms considered in Fig.~\ref{comparison_n} (b), as $N$ increases, the percentage of the total hovering time gradually increases. This trend is justified because the UAV needs more time to collect data from a larger number of ground nodes.

\section{Conclusions}\label{SecV}

In this paper, we have investigated and solved the problem of AoI-oriented data collection in UAV-enabled cluster-based IoT networks. With the aim of minimizing the total AoI of the collected data, we formulated the trajectory optimization problem as the GTSP by jointly optimizing the selection of hovering points of the UAV and the visiting order to these hovering points. To solve the formulated problem, we designed a novel algorithm framework based on the state-of-the-art transformer. In particular, the formulated trajectory planning problem is viewed as a ``translation problem''. The whole UAV-IoT network serves as the ``source language'' to the proposed model and the ``target language'' of the model is the UAV's trajectory with the minimal total AoI, where the transformer is utilized to generate the visiting order and the weighted A* is used to quickly find the hovering points. The proposed model is trained by reinforcement learning to learn a trajectory planning policy. Comprehensive experiments were conducted to evaluate the performance of the proposed algorithm. The obtained simulation results showed that the learned policy by the proposed algorithm has a strong generalization ability. When compared with other algorithms, our proposed algorithm with three different decoding methods not only reduces the total AoI, but also reduces the AoI of the oldest packet and the effective energy consumption of the UAV. Moreover, our method also has lower computation time. In future, we plan to extend the system model and the proposed algorithm to the multiple UAVs-assisted IoT network.


\appendices
\section{Proof of Lemma 1}\label{app:t1}

By substituting (\ref{plos})--(\ref{averageloss}) into (\ref{snr}), we can get the formulation (A.1) shown on top of the next page. For a fixed $H$, $20\log_{10}\left({4 \pi f_c \sqrt{H^2+R^2_{{(c_{m},b_{m})}}}} / {c}\right)$ is monotonically increasing with respect to $R_{{(c_{m},b_{m})}}$. As the analysis in \cite{al2014optimal} shows, $P_{c_{m}}^{({\rm LoS})}$ is monotonically increasing with respect to $\theta_{c_{m}}$. Since $\theta_{c_{m}} = \arctan(H/R_{(c_{m},b_{m})})$, $P_{c_{m}}^{({\rm LoS})}$ is monotonically decreasing with respect to $R_{(c_{m},b_{m})}$ for a fixed $H$. Then, ${ \left(\xi_{\text{LoS}}-\xi_{\text{NLoS}}\right)}\left/{\left(1 + \beta \exp{\left(- \widetilde{\beta}  \left(\theta_{c_{m}} -\beta\right)\right)}\right)}\right.$ is monotonically increasing with respect to $R_{(c_{m},b_{m})}$ because $\xi_{\text{LoS}} < \xi_{\text{NLoS}}$. Finally, we arrive at the conclusion that the left side of (A.1) is monotonically decreasing with respect to $R_{(c_{m},b_{m})}$ for a fixed $H$. When the SNR $\gamma_{c_{m}}$ decreases to the threshold $\gamma_{\text{th}}$, we can obtain the maximum  $R^*$. Hence, for any $c_{m}$, the UAV can successfully receive data from $b_{m}$ if $R_{(c_{m},b_{m})} \leq R^*$.

     \setcounter{equation}{0}
	\begin{figure*}[h]
	\small
		\begin{align*} \label{equ:BigWrited LiZi}
	    \nonumber
	    P_{\text{CH}}\frac{1}{\sigma^2 \left(  P_{c_{m}}^{({\rm LoS})}L_{c_{m}}^{({\rm LoS})} + \left(1-P_{c_{m}}^{({\rm LoS})}\right)L_{c_{m}}^{({\rm NLoS})} \right)} &\ge \gamma_{\text{th}}, \\ \nonumber
	     P_{\text{CH}} \frac{1}{\sigma^2 \left( 20\log_{10}\left(\frac{4 \pi f_c d_{(c_{m},b_{m})}}{v_{\text{light}}}\right) + P_{c_{m}}^{({\rm LoS})} \left( \xi_{\text{LoS}}-\xi_{\text{NLoS}}\right) + \xi_{\text{NLoS}}\right)}&\ge \gamma_{\text{th}}, \\ \nonumber
	     P_{\text{CH}} \frac{1}{\sigma^2 \left( 20\log_{10}\left(\frac{4 \pi f_c \sqrt{H^2+R^2_{{(c_{m},b_{m})}}}}{v_{\text{light}}} \right) + \frac{\xi_{\text{LoS}}-\xi_{\text{NLoS}}}{1 + \beta \exp{\left(- \widetilde{\beta}  \left(\theta_{c_{m}} -\beta\right)\right)}} + \xi_{\text{NLoS}}\right)} &\ge \gamma_{\text{th}}.
		  \tag{A.1}
		\end{align*}
		\label{total-energy}
	\end{figure*}

\end{document}